\documentclass[sigconf]{acmart}

\AtBeginDocument{%
  \providecommand\BibTeX{{%
    \normalfont B\kern-0.5em{\scshape i\kern-0.25em b}\kern-0.8em\TeX}}}

\copyrightyear{2024}
\acmYear{2024}
\setcopyright{acmlicensed}\acmConference[CHI '24]{Proceedings of the CHI Conference on Human Factors in Computing Systems}{May 11--16, 2024}{Honolulu, HI, USA}
\acmBooktitle{Proceedings of the CHI Conference on Human Factors in Computing Systems (CHI '24), May 11--16, 2024, Honolulu, HI, USA}
\acmDOI{10.1145/3613904.3642497}
\acmISBN{979-8-4007-0330-0/24/05}

\newcommand{\xhdr}[1]{\vspace{1.7mm}\noindent{{\bf #1.}}}

\newcommand{\yhdr}[1]{\vspace{1.7mm}\noindent{{\it #1.}}}

\newcommand{\nparticipants}{22\xspace}
\newcommand{\codeInterpreter}{Code Interpreter\xspace}

\usepackage{comment} 
\usepackage{xspace}
\usepackage{xcolor}

\usepackage{acmart-taps}

\def\processworkflow{procedure-oriented \behavior}

\def\processworkflows{procedure-oriented \behaviors}
\def\dataworkflows{data-oriented \behaviors}

\newcommand{\shortquote}[1]{``\emph{#1}''}
\newcommand{\longquote}[1]{\vspace{-1pt}\begin{quote}``\emph{#1}''\end{quote}}

\usepackage{xcolor}
\usepackage{colortbl}

\definecolor{lgray}{RGB}{230, 230, 230}
\definecolor{tgray}{RGB}{51, 51, 51}
\definecolor{pgray}{RGB}{217, 217, 217}
\definecolor{lteal}{RGB}{198, 219, 225}
\definecolor{tteal}{RGB}{33,90,108}
\definecolor{lblue}{RGB}{191, 225, 246}
\definecolor{tblue}{RGB}{10, 83, 168}
\definecolor{lred}{RGB}{255, 207, 201}
\definecolor{tred}{RGB}{177,2,2}
\definecolor{lorange}{RGB}{255,200,170}
\definecolor{torange}{RGB}{117,56,0}
\definecolor{lyellow}{RGB}{255,229,160}
\definecolor{tyellow}{RGB}{71,56,33}
\definecolor{lgreen}{RGB}{212,237,188}
\definecolor{tgreen}{RGB}{17,115,75}

\usepackage{soul}

\def\behavior{behavior\xspace}
\def\behaviors{behaviors\xspace}
\def\rqWorkflows{What patterns of \behaviors do analysts follow in a verification workflow?\xspace}
\def\rqArtifacts{What artifacts do analysts use and for what purposes?\xspace}
\def\rqBackground{How are analysts' backgrounds reflected in their behaviors and artifact usage?\xspace}

\usepackage{algorithm}
\usepackage{listings}
\usepackage{multirow}
\usepackage{xcolor}

\definecolor{delim}{RGB}{20,105,176}
\definecolor{numb}{RGB}{106, 109, 32}
\definecolor{string}{rgb}{0.64,0.08,0.08}

\lstdefinelanguage{json}{
    showspaces=false,
    showtabs=false,
    breaklines=true,
    postbreak=\raisebox{0ex}[0ex][0ex]{\ensuremath{\color{gray}\hookrightarrow\space}},
    breakatwhitespace=true,
    basicstyle=\ttfamily\footnotesize,
    upquote=true,
    morestring=[b]",
    stringstyle=\color{string},
    literate=
     *{0}{{{\color{numb}0}}}{1}
      {1}{{{\color{numb}1}}}{1}
      {2}{{{\color{numb}2}}}{1}
      {3}{{{\color{numb}3}}}{1}
      {4}{{{\color{numb}4}}}{1}
      {5}{{{\color{numb}5}}}{1}
      {6}{{{\color{numb}6}}}{1}
      {7}{{{\color{numb}7}}}{1}
      {8}{{{\color{numb}8}}}{1}
      {9}{{{\color{numb}9}}}{1}
      {\{}{{{\color{delim}{\{}}}}{1}
      {\}}{{{\color{delim}{\}}}}}{1}
      {[}{{{\color{delim}{[}}}}{1}
      {]}{{{\color{delim}{]}}}}{1},
}

\begin{document}

\title[]{How Do Analysts Understand and Verify AI-Assisted Data Analyses?}

\author{Ken Gu}
\orcid{0000-0002-4343-1578}
\email{kenqgu@cs.washington.edu}
\affiliation{%
  \institution{University of Washington}
  \city{Seattle}
  \state{WA}
  \country{USA}
}

\author{Ruoxi Shang}
\orcid{0000-0002-1062-5835}
\email{rxshang@uw.edu}
\affiliation{%
  \institution{University of Washington}
  \city{Seattle}
  \state{WA}
  \country{USA}
}

\author{Tim Althoff}
\orcid{0000-0003-4793-2289}
\email{althoff@cs.washington.edu}
\affiliation{%
  \institution{University of Washington}
  \city{Seattle}
  \state{WA}
  \country{USA}
}

\author{Chenglong Wang}
\orcid{0000-0002-5933-6620}
\email{chenwang@microsoft.com}
\affiliation{%
  \institution{Microsoft Research}
  \city{Redmond}
  \state{WA}
  \country{USA}
}

\author{Steven M. Drucker}
\orcid{0000-0002-5022-9343}
\email{sdrucker@microsoft.com}
\affiliation{%
  \institution{Microsoft Research}
  \city{Redmond}
  \state{WA}
  \country{USA}
}

\renewcommand{\shortauthors}{Gu et al.}

\begin{abstract}
Data analysis is challenging as it requires synthesizing domain knowledge, statistical expertise, and programming skills. Assistants powered by large language models (LLMs), such as ChatGPT, can assist analysts by translating natural language instructions into code. However, AI-assistant responses and analysis code can be misaligned with the analyst's intent or be seemingly correct but lead to incorrect conclusions. Therefore, validating AI assistance is crucial and challenging. Here, we explore how analysts understand and verify the correctness of AI-generated analyses. To observe analysts in diverse verification approaches, we develop a design probe equipped with natural language explanations, code, visualizations, and interactive data tables with common data operations. Through a qualitative user study (n=22) using this probe, we uncover common behaviors within verification workflows and how analysts' programming, analysis, and tool backgrounds reflect these behaviors. Additionally, we provide recommendations for analysts and highlight opportunities for designers to improve future AI-assistant experiences.
\end{abstract}



\maketitle

\begin{figure*}[h!]
  \centering
  \includegraphics[width=0.92\linewidth]{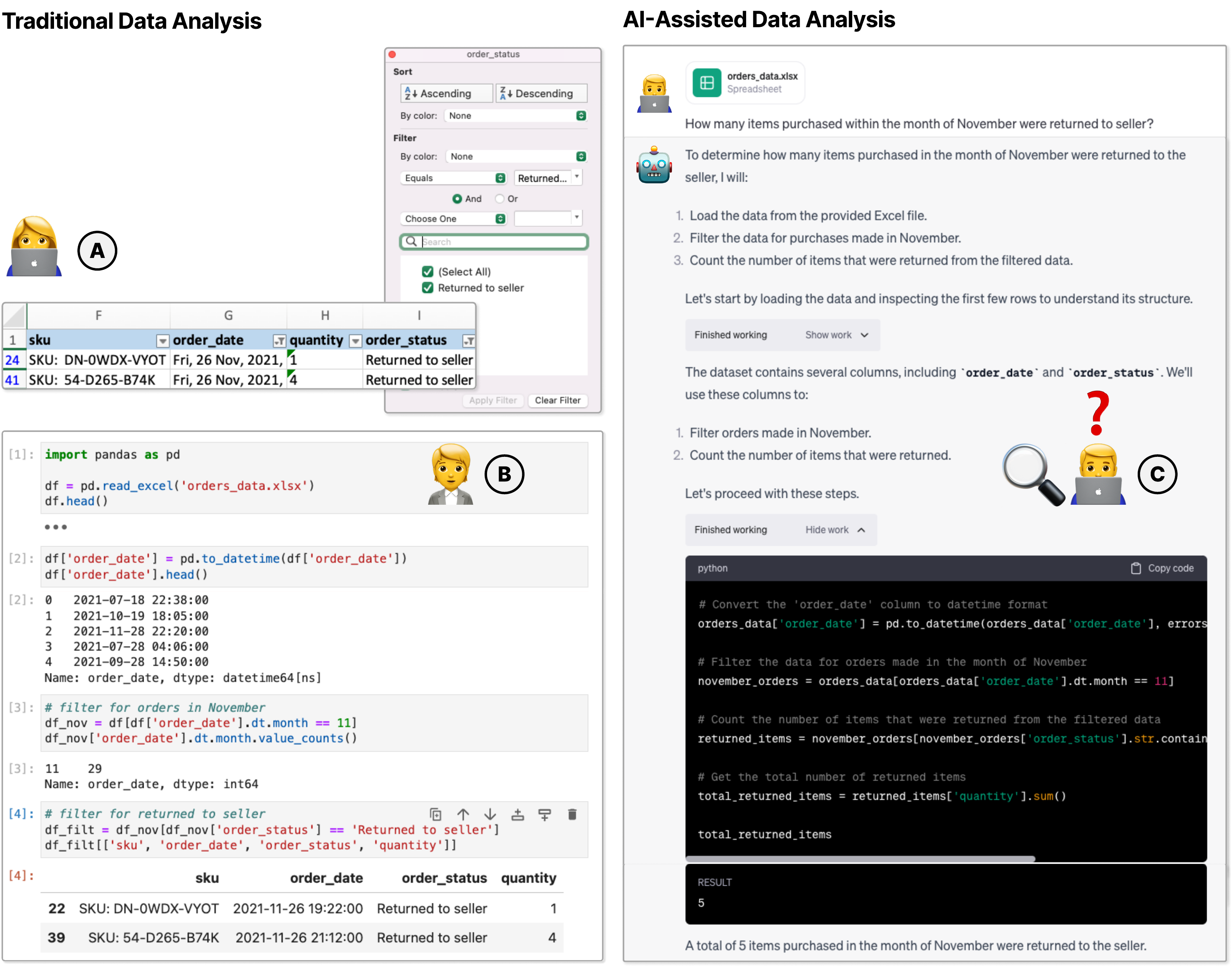}
  
  \caption{\textbf{Analysts may now need to understand and verify AI-assisted analyses}. In traditional analysis workflows, analysts specify and execute their data operations using tools such as computational notebooks (A) or spreadsheets (B). Engaged in these operations, analysts are familiar with the process and results of their work (e.g., reports, code, tables, and visualizations). However, with AI-assisted analysis, analysts can convey their intentions using natural language (e.g., "\textit{How many items purchased within the month of November were returned to the seller?}"). The AI assistant handles the task of specifying and performing the data operations. This shift requires analysts, including those who may not be familiar with the underlying execution language the assistant uses, to understand and verify the process and results of the assistant (C). In this paper, we study the workflows analysts with varied backgrounds use to understand and verify AI-assisted data analyses. 
  }
  \Description{Three images labelled A, B, and C showing the screenshots of the the data analysis tools. A and B are on the left side under the title "Traditional Data Analysis". C is on the right side with the title "AI-Assisted Data Analysis". Image A shows the partial interface of working with spreadsheets. Image B shows the partial interface of a Jupyter notebook during data analysis. Image C shows the chatbot interface between a user and an AI assistant where the user asks a question and the AI produces the data analysis output. All three images are labelled with emojis of different people.}
  \label{fig:motivate}
\end{figure*}

\section{Introduction}
\label{sec:intro}
Data analysis is challenging as practitioners must synthesize domain knowledge with computational expertise to draw reliable conclusions. For practitioners, computational challenges may act as a barrier to exploring alternative and valid analysis decisions, and therefore conducting robust analyses~\cite{Liu2020UnderstandingTR, Liu2019PathsEP}.

Improved capabilities of large language models (LLMs) for general~\cite{Chen2021EvaluatingLL, Bubeck2023SparksOA, Chowdhery2022PaLMSL}, data science ~\cite{Lai2022DS1000AN, Yin2022NaturalLT, Chandel2022TrainingAE}, and visualization~\cite{Dibia2023LIDAAT} programming tasks suggest LLM-based analysis assistants as a promising solution, enabling data analysts to \textit{execute} and \textit{automate} their analyses. Given the widespread deployment of assistants such as OpenAI's \codeInterpreter ~\cite{openai_codeinterpreter}, there is an emerging paradigm in how analysts perform analyses. Rather than formulate the exact transformations and operations in an analysis, analysts specify their analysis question through a natural language intent~(Fig.~\ref{fig:motivate} right)~\cite{Liu2023WhatIW}. The AI assistant then performs the analysis task, writes code to understand the data, makes the necessary data transformations, and interprets the results. Since many analysis tasks are time-consuming~\cite{Chattopadhyay2020WhatsWW, Kandogan2014FromDT, Kandel2012EnterpriseDA, Liu2020UnderstandingTR},  even for analysts well-versed in programming, natural language interactions can be more efficient especially when abstraction of the intent is possible. Furthermore, this natural language communication can broaden the usability of AI analysis tools, making them accessible even for analysts who are non-expert programmers~\cite{Ragavan2022GridBookNL}.

Yet, a significant challenge remains. AI-based assistants, while powerful, can misinterpret user intentions (see Fig.~\ref{fig:motivate})
~\cite{Liang2023UnderstandingTU, Xu2021InIDECG, Jiang2022DiscoveringTS} or make seemingly correct erroneous outputs~\cite{Bender2021OnTD, Ji2022SurveyOH}. Since misinterpretations is inherent in the flexibility of natural language communication~\cite{Liu2023WhatIW, Ragavan2022GridBookNL, Setlur2022HowDY}, errors from existing AI-based assistants are likely to persist no matter what the underlying LLM is. Moreover, because conclusions from data analyses often inform high-impact decisions in science~\cite{Baker20161500SL, Aarts2015EstimatingTR}, business~\cite{Kandel2012EnterpriseDA, Kim2018DataSI, Kim2016TheER}, and government~\cite{Hardy2017OpeningUG, Ubaldi2013OpenGD}, it is important for users of AI-based analysis assistants to critically reflect on the assistant’s outputs~\cite{Ragavan2022GridBookNL}. In this new paradigm, analysts' focus now shifts from writing and validating their own analyses to understanding and evaluating the analyses generated by AI assistants~\cite{Liang2023UnderstandingTU}. This can be challenging as existing AI assistants lack verification support~\cite{Ragavan2022GridBookNL, Sarkar2022WhatII} and even experts can be susceptible to automation bias and overreliance~\cite{Wickens2015ComplacencyAA, Parasuraman2010ComplacencyAB}.

Recent work in the HCI community explored ways to provide the user with a better understanding of LLM-generated outputs, including those from AI-based code assistants~\cite{Sun2022InvestigatingEO, Vasconcelos2023GenerationPA}. However, these studies focus only on programmers and not the specific domain of data analysis, missing a significant portion of data analysts~\cite{Crisan2020PassingTD, Liu2020UnderstandingTR}. Since data analysis involves iteratively making sense of the underlying data~\cite{Koesten2019TalkingDU, Pirolli2007TheSP, Klein2007ADT}, analysts must juggle their mental model of the data in addition to their understanding of the AI's procedure. There is limited insight into how analysts with different programming skills and computational reasoning abilities go about this process. 
Current code assistants such as Open AI's Code Interpreter~\cite{openai_codeinterpreter} present these steps in natural language, Python code, code comments, and code execution outputs. It is unclear to what extent these interface affordances are helpful and how analysts use these different sources of information for accessing the correctness of AI-generated analyses. 

In this work, we build on the perspective of recent works in Human-Centered Explainable AI (HCXAI)~\cite{Ehsan2021ExplainabilityPB, Liao2020QuestioningTA, Liao2021HumanCenteredEA, Kim2022HelpMH} in that users' needs to understand AI-generated outputs are dependent on their backgrounds, social, and organizational contexts; we focus on data analysts as they understand AI-assisted analyses. We study the verification workflows of analysts, how different backgrounds may influence analysts' approaches, and opportunities to improve this experience. We aim to guide the design of AI assistants that empower analysts to critically reflect on AI-generated outputs. 

To explore analysts' behaviors and elicit feedback for design opportunities, we developed a prototype interface that not only includes the AI's code, code comments, and natural language explanation but also additional interactive data tables and summary visualizations (Sec.~\ref{sec:method}). This design accommodated a broad spectrum of potential artifacts that analysts may employ during verification, facilitating the observation of naturally emerging workflows.
Analysts could access intermediary data tables that let them filter, search, and sort for values to explore the data and verify the AI-generated output. Using our tool as a probe, we conducted a qualitative study with \nparticipants professional analysts (Sec.~\ref{sec:study}). We asked participants to verify AI-generated analyses that involved sequences of data transformations covering a range of real-world datasets (Table~\ref{tab:tasks}). These generated analyses were in response to high-level analysis queries written by data scientists~\cite{Yin2022NaturalLT} which were presented to participants as prompts written by an intern.

We find analysts often started their verification workflows with \textit{procedure-oriented} behaviors (answering \shortquote{what did the AI do?}) and shifted to \textit{data-oriented} behaviors (answering \shortquote{does the [resulting] data make sense?}) once they noticed issues in the AI-generated output (Table~\ref{tab:freq-of-workflows}). Analysts adopted procedure-oriented behaviors to get a high-level understanding of the analysis steps and confirm low-level details in the data operations. Meanwhile, analysts adopted data-oriented behaviors to make sense of the data. 
Notably, \textit{data artifacts} (i.e., the data tables and summary visualizations) were often vital as secondary support in \textit{procedure-oriented} \behaviors, and vice versa for \textit{procedure artifacts} (i.e., the natural language explanation and code/code comments) (Fig.~\ref{fig:artifacts}). 

Based on these findings (Sec.~\ref{sec:results}), we make recommendations to end-user analysts interacting with AI-based analysis assistants (Sec.~\ref{sec:discussion_impl_analyst}). Additionally, we discuss  implications for system designers (Sec.~\ref{sec:discussion_impl_designers}). These center around fluidly connecting data-oriented and procedure-oriented artifacts, communicating data operations and AI's assumptions about the data, and incorporating AI guidance into verification workflows.

This paper contributes the following:
\begin{enumerate}

\item Findings from a user study using our design probe that uncover common behavioral patterns in the verification of AI-generated analyses,
\item A set of implications for end-user analysts interested in AI-assisted data analysis, paired with design implications for tool builders to improve verification workflows.
\end{enumerate}

\section{Motivating Example}
To illustrate the potential errors and verification workflows in AI-assisted analyses, we compare three hypothetical analysts working on the same analysis question~(Fig.~\ref{fig:motivate}). This scenario highlights the workflow differences between traditional and AI-assisted data analysis. Jane, Kate, and Alex are data analysts at a company that sells products on a popular e-commerce platform. They want to assess how many products may have quality control issues as the year draws to a close. Specifically, they want to find out how many distinct products bought in November were later returned.
To perform this analysis, they have a dataset of product orders where each row is a unique order for a product and its associated order status. The dataset contains the following key columns:
\begin{itemize} 
  \item \textbf{order\_date}: the date which the order was placed
  \item \textbf{sku}: a unique identified for the product
  \item \textbf{quantity}: the number of units of the product ordered
  \item \textbf{order\_status}: indicating whether the order was \textit{Delivered to Buyer} or \textit{Returned to Seller}
\end{itemize}


\xhdr{Traditional Data Analysis.} Jane uses a spreadsheet for her analysis (Fig.~\ref{fig:motivate}A). She explores the dataset in her spreadsheet tool. After grasping its structure, she zeros in on the task. She uses the spreadsheet's interface to filter the \textbf{order\_date} column for the month of November and the \textbf{order\_status} column for \textit{Delivered to Buyer}. She sees that there are two rows left, indicating two unique products were ordered in November and returned to the seller.  

In contrast, Kate opts to use computational notebooks for data analysis, employing Python code to investigate the analysis question (Fig.~\ref{fig:motivate}B). She writes code to read the dataset and to understand the data. 
She then writes code to filter the \textbf{order\_date} column for the month of November and the \textbf{order\_status} column for \textit{Delivered to Buyer}. In her workflow, Kate validates her operations by writing code to get a quick sense of the intermediate data produced. Like Jane, Kate arrives at the same conclusion. Throughout this process, Jane and Kate are actively engaged in the data and operations, giving them confidence in the correctness of their conclusion.

\xhdr{AI-Assisted Data Analysis}
Alex, meanwhile, decides to leverage an AI assistant for help. He uploads the dataset and writes a natural language prompt: \textit{How many items purchased within the month of November were returned to seller?}

The AI assistant writes and executes code to read the dataset. Next, it performs the analysis task, generating and running code before arriving at a final answer of 5. During this, the AI assistant also explains its steps in natural language. Because Alex was not directly involved in the analysis, he needs to validate its answer by inspecting the AI-generated output (Fig.~\ref{fig:motivate}C). He reads over the AI's natural language explanation of its procedure and skims the code briefly. Without fully understanding the code, he deems the answer to be correct. However, upon comparing his answer with Jane's and Kate's, Alex realizes the AI made a mistake. Looking through the intermediate data in Kate's work, Alex realizes the AI assistant calculated the number of units returned rather than the number of \textit{unique} products returned. 

The errors in AI-assisted data analysis are subtle and require the analyst to carefully scrutinize the AI's procedure and associated data involved. Additionally, current AI assistants do not facilitate the exploration of data along with the AI's code and natural language explanation, thereby constricting analysts' verification workflows. Further, it is unclear how analysts go about verification. Therefore, this paper seeks to understand analysts' workflows and identify opportunities for improvement.

\section{Background and Related Work}

\subsection{AI-Based Tools for Data Science Execution}
\label{sec:bg_automated}

AI-based code assistants can enable analysts to conduct analyses as end-user programmers~\cite{Liu2023WhatIW, Ko2011TheSO}. With tools such as Github Copilot and ChatGPT, analysts' intent specified through natural language can be more expressive and approachable than those conveyed in domain-specific programming languages~\cite{DeLine2021GlindaSD, Gulwani2014NLyzeIP}. Given the potential of AI-supported natural language programming, there is a growing body of research in the value and usability of these tools~\cite{Sarkar2022WhatII, Liang2023UnderstandingTU, Vaithilingam2022ExpectationVE, Baker20161500SL, Dibia2022AligningOM, GitHubCopilotResearch}. These works focus on programmer experiences with AI-based code assistants and find common challenges in comprehending and verifying the AI-generated code~\cite{Liang2023UnderstandingTU, Vaithilingam2022ExpectationVE, Sarkar2022WhatII}. 

Here, we build on prior work and study the verification challenges for data analysts as end-user programmers~\cite{Ko2011TheSO}. Compared to software engineering, data analysis is more exploratory and iterative~\cite{Kery2017ExploringEP, Pirolli2007TheSP, Liu2019PathsEP, Rule2018ExplorationAE} as code is not the deliverable but the outcome of the analysis~\cite{Epperson2022StrategiesFR, Kandel2012EnterpriseDA}. Importantly, making sense of the data is an integral part of data analysis~\cite{Pirolli2007TheSP, Russell1993TheCS, Grolemund2014ACI, Koesten2019TalkingDU}. While existing work has studied how programmers validate AI-written code for general programming tasks~\cite{Barke2022GroundedCH}, our work observes how data analysts' \textit{sensemaking} behaviors interplay with their verification of AI-generated analyses.

AI-based analysis assistants have also received great attention. Prior work has explored the design of analytical chatbots~\cite{Zhi2020GameBotAV, Kassel2018VallettoAM, Hoon2019InterfacingCW} such as what should the responses be from such an assistant~\cite{Setlur2022HowDY} and how analysts can request, specify, and refine assistance~\cite{Mcnutt2023OnTD}. Given the attention and promise, these assistants have been integrated into popular analysis environments such as computational notebooks~\cite{PerryMallick2023, Gu2023HowDD} and spreadsheets~\cite{Stallbaumer2023}, and released as standalone systems~\cite{Dibia2023LIDAAT, Datagran2023}. In contrast, we examine these tools from the perspective of supporting analysts' understanding and verification of AI-generated analyses, especially as AI-assistants can specify sequences of data operations (e.g., Fig.~\ref{fig:motivate} right). In particular, we contribute the first user study examining the common behaviors and challenges in verification workflows.

\subsection{Understanding and Sensemaking in Data Analysis}
\label{sec:bg_understand_analysis}
Prior work has found that data scientists often compare conclusions with prior analyses and existing data~\cite{Kim2016TheER, Liu2019PathsEP}. Likewise, as part of data science collaboration and communication, it is common for data analysts to interpret and understand the results of another analyst's work~\cite{Zhang2020HowDD, Pang2022HowDD, Crisan2020PassingTD, Kim2018DataSI}. During this sensemaking, analysts often need help managing, tracking, and comparing the iterative and messy nature of their computations~\cite{Kery2019TowardsEF, Kery2018TheSI, Head2019ManagingMI, Wang2020AssessingAR}. To support this, prior work has explored how to help analysts track the data provenance in their analyses~\cite{Kery2019TowardsEF, Kery2017VarioliteSE, Wu2020B2BC} and visualize how their data changes throughout various data operations~\cite{Wang2019HowDS, Pu2021DatamationsAE, Xiong2022VisualizingTS}.

While existing research has focused on understanding human-written analyses, there are key differences between understanding analyses written by a human, and those written by an AI. People develop diverse perceptions and levels of trust in AI systems~\cite{Khadpe2020ConceptualMI, Jung2022GreatCO, Devito2018HowPF, Gu2023HowDD}, leading to distinct behaviors when evaluating AI-generated analyses. For example, McNutt et al.~\cite{Mcnutt2023OnTD} observed that some data scientists naturally trust analyses written by a colleague and may overlook code auditing. Additionally, social factors in data analysis communication often result in communicating asynchronously~\cite{Wang2019HowDS}, employing multiple platforms (i.e., Slack, Zoom, Google Docs etc.)~\cite{Pang2022HowDD}, and selectively sharing information~\cite{Pang2022HowDD, Liu2019PathsEP}. These qualities are less likely to appear when interacting with an AI-assistant.
Therefore, AI analysis assistants present a unique paradigm for analysis verification and sensemaking. Given the widespread deployment of AI-based tools, our work seeks to understand the intricacies of this paradigm.

\subsection{Understanding and Verifying LLM Outputs}
\label{sec:bg_verify_ai}

Prior work in the Human-Centered Explainable AI community has emphasized the goal that understanding AI revolves around empowering individuals to fulfill their objectives~\cite{Ehsan2021ExplainabilityPB, Liao2020QuestioningTA, Liao2021HumanCenteredEA, Kim2022HelpMH}. Following this perspective, recent work has explored ways to make LLMs more interpretable as they present their own explainability challenges~\cite{Sun2022InvestigatingEO, Vasconcelos2023GenerationPA, Dibia2022AligningOM}. Specifically, these works have explored opportunities for AI-code assistants in programming applications and proposed different explainable AI (XAI) features to guide the programmer's understanding of the model. In contrast, our work focuses on XAI features for data analysts. As data sensemaking is inherent in data analysis~\cite{Grolemund2014ACI, Koesten2019TalkingDU}, we explore specific features that help analysts in following the AI's data operations (e.g., via the natural language explanation and associated data tables).

We note that understanding and verifying AI-generated outputs can involve additional follow-up prompts to the AI assistant~\cite{Liu2023WhatIW, Lakkaraju2022RethinkingEA}. In this work, we focus on understanding the usage of supporting artifacts and disentangle the verification workflow using these artifacts from prompt writing (Sec.~\ref{sec:method_design}). Works studying abstraction matching~\cite{Liu2023WhatIW}, prompt writing~\cite{ZamfirescuPereira2023WhyJC}, and repair of AI-generated programs~\cite{Olausson2023DemystifyingGS, Ashktorab2019ResilientCR} are related but distinct.

\section{Method: Design Probe and Prepared Tasks}
\label{sec:method}

In this work, we aim to study how analysts verify AI-generated analyses with features that support data sensemaking. In doing so, we wanted to minimize the impact of variations in the dataset, task difficultly, prompt expertise, and AI responses. 
To facilitate analysts' natural sensemaking workflows when working with data, we chose to purpose-build a design probe~\cite{boehner2007hci} and provide additional data artifacts based on recommendations from prior work~\cite{Alspaugh2019FutzingAM, Kery2019TowardsEF, Choi2023TowardsTR, Batch2018TheIV, Kery2018InteractionsFU, Wu2020B2BC, Moritz2017TrustBV}. We made this decision because existing assistants, such as Code Interpreter~\cite{openai_codeinterpreter}, do not allow easy exploration of data and only provide natural language explanations, code, and execution outputs. 
In this section, we discuss the key design considerations involved in our study (Sec.~\ref{sec:method_design}) before outlining the specific design of our probe (Sec.~\ref{sec:method_probe}) and the preparation of our study tasks (Sec.~\ref{sec:method_materials}).

\subsection{Design Considerations}
\label{sec:method_design}
In building our design probe and formulating our user study, we identified three central design considerations that involved different trade-offs outlined below. These decisions are guided by our goal to make reliable observations on analysts' \emph{verification workflows} with different \emph{supporting artifacts}, while minimizing distractions from other aspects of the data analysis workflow.

\xhdr{Using familiar and approachable supporting artifacts} 
A crucial decision in implementing our design probe revolved around the choice of data artifacts. One option is to incorporate customized feature-rich artifacts (e.g., those in \textit{Datamations} or \textit{DITL}~\cite{Wang2022DiffIT, Pu2021DatamationsAE}) and verification interfaces (e.g., incorporating visualization tools like \textit{Voyager}~\cite{Wongsuphasawat2016VoyagerEA, Wongsuphasawat2017Voyager2A} or \textit{PygWalker}~\cite{kanaries_pygwalker} for verificaiton). Alternatively, we could opt for integrating a core set of simpler, more user-friendly, and approachable data artifacts. These two options contrast with each other in terms of \emph{the richness of actions analysts can achieve with the interface} and \emph{the time and effort analysts need to spend to learn to perform these actions}. 

Ultimately, we opted for familiar and approachable artifacts as our study focus is to observe verification workflows: we determined that the challenges tied to learning and interpreting customized visual elements would limit our ability to observe these workflows~\cite{Grammel2010HowIV, Kwon2011VisualAR, Lee2016HowDP}. Likewise, while generic artifacts and interfaces may overlook some features experienced analysts could leverage, they let us focus on analyzing analysts' workflows across a broad audience and still gather concrete feedback on the features they want.

\xhdr{Focus on one turn of the human-AI interaction} 
Another choice that shaped our probe and study was whether or not to allow follow-up prompts to the AI as part of the verification workflow. On one hand, choosing to allow follow-up prompts enables analysts to use them in their verification. However, besides interpreting the AI's current output, analysts would need to grapple with the challenge of formulating prompts~\cite{ZamfirescuPereira2023WhyJC, Mishra2022HELPMT}. On the other hand, opting against follow-up prompts mitigates the impact of prompt expertise but at the cost of confining analysts' actions. In our study, we chose to zoom in on one turn of the human-AI interaction and disallow follow-up prompts. As every new prompt triggers a corresponding response from the AI assistant, continuous interpretation and verification of AI-generated outputs are inevitable. In reducing the scope, we chose to thoroughly observe how analysts understand the AI's outputs and interact with supporting artifacts. Similarly, aiding analysts in understanding the AI-generated output through artifacts  not only helps them rectify errors in immediate interactions but also craft more effective prompts in future engagements.

\begin{figure*}[t]
  \centering
  \includegraphics[width=1.0\linewidth]{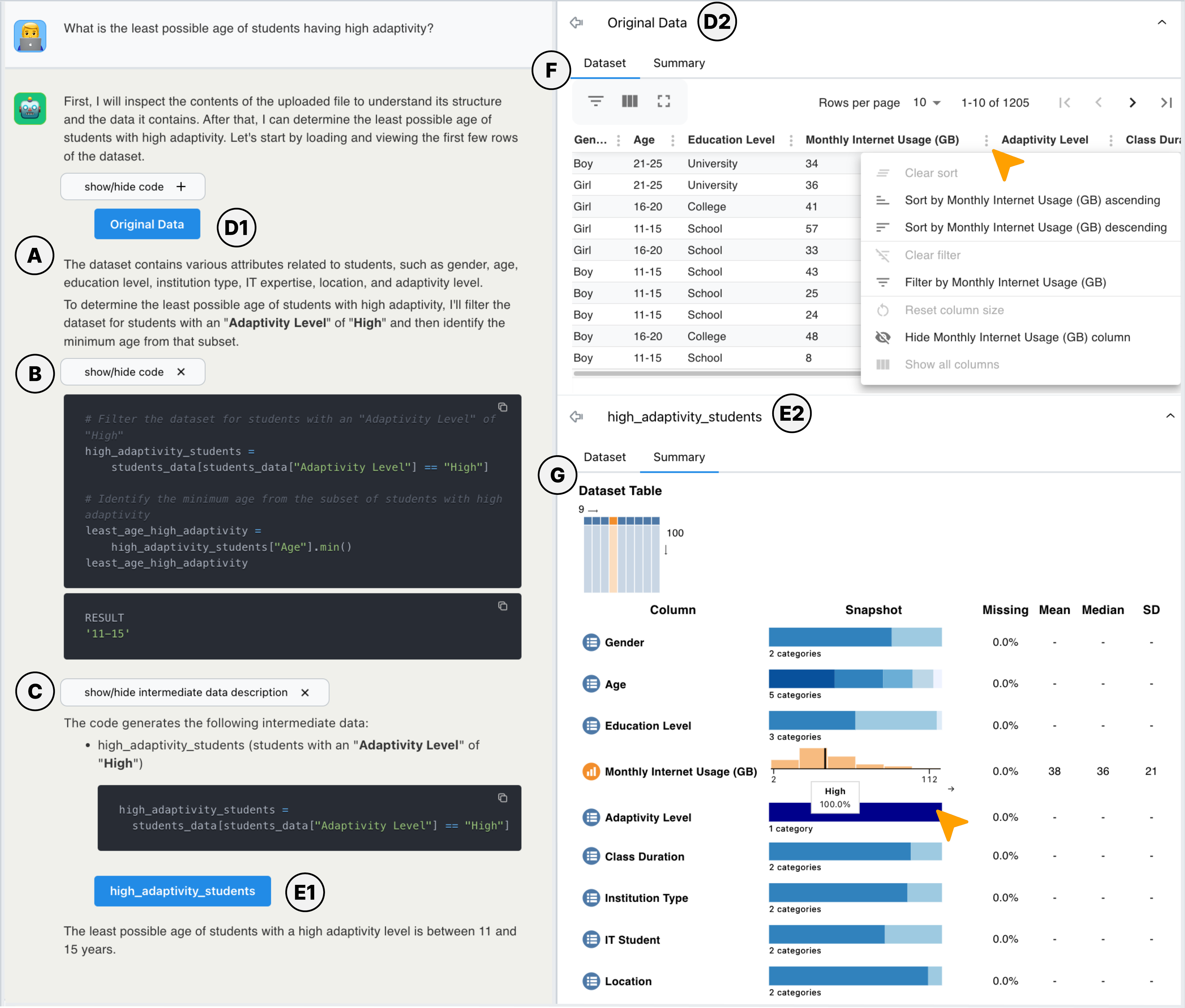}
  \caption{\textbf{Probe Interface}. The analyst's prompt and the assistant's response are shown in the left panel. The AI's response includes its natural language explanation (A), the code and code comments involved in its calculations (B), and a description of any intermediate data (C). The original data table and intermediate data table(s) are accessible via buttons interleaved in the AI's response (D1 and E1) and when clicked point to their corresponding data pane in the right panel (D2 and E2). These panes can also be opened directly in the right panel. In each pane, analysts can view the raw data table in the \textit{Dataset} tab with sort and filter functionality (F). Analysts can also view a visualization showing the distribution and basic descriptive statistics of each column in the \textit{Summary} tab (G). }
  \Description{The screenshot of the Probe Interface web page with double column design. The page consists of two large main content areas with the chatbot interface on the left and the data tables displayed on the right. The screenshot is labeled with letters A to G that are used to describe the different features in the figure caption.}
  \label{fig:interface}
\end{figure*}

\xhdr{Prepare standardized task materials} The choice of datasets and verification tasks was also important to consider. One option is to allow analysts to work with their own datasets while interacting with the AI in real-time to verify the AI's outputs. Alternatively, we could pre-prepare the natural language prompts, associated datasets, and AI-generated responses. These choices underscore a trade-off between matching analysts' natural workflows and mitigating variability associated with generative AI and dataset complexity. In our study, to control potential confounding factors, we chose to prepare all task materials. Although this approach may differ from the scenario where analysts' verify their own datasets and analyses, it closely aligns with situations where analysts need to familiarize themselves with another analyst's analysis~\cite{Pang2022HowDD, Zhang2020HowDD, Koesten2019CollaborativePW}. To mitigate concerns related to dataset unfamiliarity, we carefully prepared approachable and realistic datasets from diverse domains, used prompts written by real-world data scientists, and provided additional background descriptions of the dataset and analysis motivation (Sec.~\ref{sec:method_materials}). Given our chosen approach, it would be interesting to apply the lessons learned in this study to analysts working on their own datasets and tasks (Sec.~\ref{sec:limitation}).

\subsection{Design of the Probe}
\label{sec:method_probe}
Data analysts need to \textit{inspect} the data (i.e., the data quality, format, columns, values etc.) and \textit{engage} with the data (e.g., understanding relationships between columns, looking for outliers etc.) as part of data sensemaking activities~\cite{Koesten2019TalkingDU}. Therefore, we designed our probe with facilities for understanding the data. This includes data tables, and summaries and histograms of all the fields in the data. To support sensemaking while making components approachable, our probe incorporates abstractions of modalities found in popular data science tools: the code, code comments, natural language explanations, data tables, and visualizations~\cite{jupyterlab, tableau, rstudio, Drosos2020WrexAU, Kery2020mageFM, diamondreivich2020mito, Wu2020B2BC}.

\xhdr{User Interface}
The probe's interface follows a multi-panel design. The main panel shows the chat interface between the user and the assistant (Fig.~\ref{fig:interface} left). The user's prompt and subsequent response from the assistant are shown. In the assistant's response, following Code Interpreter~\cite{openai_codeinterpreter} and explanation practices recommended and found favorable in prior work~\cite{Lakkaraju2022RethinkingEA, Sun2022InvestigatingEO, Drosos2020WrexAU}, the interface shows the assistant's natural language explanation of its procedure, associated code, code comments, and execution outputs (Fig.~\ref{fig:interface}A and B). These are interleaved together such that all artifacts are available at any time. In addition, because understanding the sequence of data manipulations is vital to understanding the output~\cite{Pu2021DatamationsAE, Xiong2022VisualizingTS},  the probe includes descriptions of any intermediate data computed as a result of an atomic data wrangling operation (Fig.~\ref{fig:interface}C)~\cite{Xiong2022VisualizingTS, Kasica2020TableSA}.

 Inspired by prior work detailing the importance of contextualized intermediate data artifacts for analysis provenance~\cite{Alspaugh2019FutzingAM, Kery2019TowardsEF, Choi2023TowardsTR, Batch2018TheIV, Kery2018InteractionsFU, Wu2020B2BC, Moritz2017TrustBV}, the probe includes a side data panel (Fig.~\ref{fig:interface} right) that allows analysts to view all relevant data in the AI-generated analysis. Having this as a side panel is helpful as it suggests the information is of global scope~\cite{Mcnutt2023OnTD}. Additionally, by keeping the data separate from the AI-generated analysis, the probe minimizes any interference to analysts following the AI's procedure in the main panel. In the side panel, data tables are shown (i.e., the original data and the results of intermediate calculations) as collapsible panels. Each collapsible panel includes a raw data table and a summary visualization of the table from ~\cite{observablehq_summary_table}. The summary visualization offers a quick visual summary of the overall table, the values and distribution of each column, and basic descriptive statistics.

\xhdr{User Interactions}
In the main panel, the probe's interactions prioritize easy access to the full/high-level abstraction of the procedure conveyed through the AI's natural language explanation. Because code and table descriptions can obstruct and interrupt this view, the code and intermediate data descriptions in the main panel are initially hidden which users can reveal via a show/hide button. Analysts can also access the side data panels within the flow of the AI's explanation via in-context buttons which open and close the corresponding panel (Fig.~\ref{fig:interface}E1 and E2).

In the side panel, analysts can perform basic operations on each table such as sorting columns and filtering rows based on specific values in the column(s). To support easy navigation between the data tables and AI's procedure in the main panel~\cite{Choi2023TowardsTR, Wu2020B2BC}, clicking on a data table scrolls the main panel to the corresponding context of that data table.

\begin{table*}[t]
    \small
 
  \begin{tabular}{lp{1.1cm}p{5.7cm}p{6.0cm}p{1.6cm}}
    \toprule
    \textbf{TID} & \textbf{Task} & \textbf{Natural Language Prompt} &  \textbf{Error Type and Summary}&\textbf{PIDs} \\
    \midrule
    T0~\cite{Tutorial} & \raggedright Tutorial & What is the least possible age of students having high adaptivity?  & \textit{Data}---AI and took the minimum of age-range by alphanumeric order.  &  Everyone \\
    \midrule
    T1~\cite{AmazonOrders} & \raggedright Amazon Orders & How many items purchased within the month of November were returned to seller?  & \textit{Prompt}---AI counted the quantity of products returned, instead of the number of unique products.&  P1, P7, P15, P20 \\ 
    \midrule
    T2~\cite{BigBasket} & \raggedright Big Basket1& Show a list of the top five rated Nivea products. & \textit{Data}---AI missed "Nivea Men" products in the data. &  P1, P2, P5, P12 \\
    \midrule
    T3~\cite{BigBasket} & \raggedright Big Basket2&  How expensive are gourmet products compared to beverage products in average? Show the value as a percentage of beverage products. & \textit{Correct} &    P1, P2, P5, P12, P22 \\
    \midrule
    T4~\cite{Bollywood} & \raggedright Bollywood & Show the top 5 movies with the highest percentage return on investment. & \textit{Calculation}---AI mentioned but did not ignore rows with bad budget values. The intern did not filter for only Bollywood movies. &  P2, P6, P10, P15, P21  \\
    \midrule
    T5~\cite{Flights} & \raggedright Flights1& What is the city that has the highest number of incoming flights? How many incoming flights does each airline have for that city with one week left until departure? Show the city name, airline and number of incoming flights. & \textit{Calculation \& Data}---AI performed the wrong order of filtering. It also incorrectly counted incoming flights based on booking rows rather than by unique flight codes.  &  P4, P8, P10, P14, P17\\
    \midrule
    T6~\cite{Flights} & \raggedright Flights2&  What is the least expensive time of the day to depart to Chennai in Economy class across all airlines? Show the time of the day and the fare price for each airline. &\textit{Prompt}---For each airline, the AI calculated the minimum price of all flights during a time of day rather than the minimum of the average price of flights during a time of day. &  P4, P8, P10, P14, P17, P20 \\
    \midrule
    T7~\cite{Hotels} & \raggedright Hotels& Which hotels had a worse ranking this year than in 2021? Show the hotel name, location and the difference in ranking from last year. & \textit{Data}---AI misinterpreted a "2021" flag column as representing the rank in 2021. &  P3, P4, P6, P8, P12, P16  \\
    \midrule
    T8~\cite{MovieContent} & \raggedright Movie Content& What is the number of shows viewed in distinct languages within each genre as a percentage of the total number of shows within each genre? Show the genres as an index and languages as columns & \textit{Calculation}---Inverted the percentage calculation. Summing across genres for a given language should add up to 100 but the AI's calculation resulted in summing across languages for a genre adding up to 100. &  P5, P7, P11, P19, P22 \\
    \midrule
    T9~\cite{Netflix} & \raggedright Netflix& Who is the actor who worked with the same director the most? &  \textit{Correct} &   P6, P9-11, P13, P18, P19, P21  \\
    \midrule
    T10~\cite{TV} & \raggedright TV&What are the top 5 most selling television frequencies? Show the frequencies with their counts &  \textit{Data}---The AI cleaned dirty frequency data but it neglected correct data that existed in other columns. &  P7, P13, P20, P22 \\
    \bottomrule
  \end{tabular}
  \caption{\textbf{User Study Tasks.} These tasks and natural language prompts were derived from Yin et al.~\cite{Yin2022NaturalLT}. The prompts were written by data scientist annotators, who were instructed
  to be natural, concise, and avoid unnecessary elaborations. AI errors encompassed poor interpretations of the prompt and data, and incorrect calculations. Although creating more precise prompts can potentially reduce errors, our study is not centered on this aspect. 
  }
   \label{tab:tasks}
\end{table*}

\subsection{Preparation of Study Tasks}
\label{sec:method_materials}
To prepare realistic analysis verification tasks and AI-written errors representative of those made by state-of-the-art AI assistants, we used the ARCADE benchmark dataset from Yin et al.~\cite{Yin2022NaturalLT}. ARCADE features multiple rounds of natural language intent to code problems (i.e. a natural language query, dataset, and code solution) situated in computational notebooks\footnote{these notebooks used datasets uploaded to \href{https://www.kaggle.com}{Kaggle} and encompassed data wrangling and EDA operations} written by data scientist annotators. In our task preparation, we ran queries from ARCADE on Code Interpreter~\cite{openai_codeinterpreter} and carefully selected a subset of queries that contained errors, had good coverage across a range of domains and error types, and were approachable to analysts (details in the appendix). These queries and Code Interpreter responses formed the backbone of our tasks. 

 In total, we prepared one tutorial task and 10 main tasks, 2 of which did not contain any errors, spanning 9 unique datasets (Table~\ref{tab:tasks}). For each task, we used the generated code from Code Interpreter's output to prepare intermediate data tables and table summary visualizations. To provide analysts with sufficient information, we prepared additional intermediate data tables for all major transformations, often breaking up chained function calls in the code. We then verified with multiple expert data scientists that these intermediate data tables were sufficient for the task. To ease dataset understanding and help participants focus on verification, we reduced the number of columns for excessively complex datasets. We made sure to leave enough extra columns irrelevant to the immediate analysis query such that the task remains non-trivial. In addition, we provided a clear ground truth motivation and a more detailed description of the analysis query. In the rest of the paper, to differentiate the two, we refer to the detailed description of the analysis query as the \textit{analysis goal}. Finally, we
provided a description of the dataset and what each column in the dataset represented.

In summary, we compiled the following primary materials for each task: the natural language query from ARCADE, the unmodified code, code comments, and natural language explanation response from Code Interpreter, the prepared data tables and visualizations (Fig.~\ref{fig:interface} right), and a detailed description of the motivation, analysis goal, and dataset.


\section{User Study}
\label{sec:study}
\begin{table*}
    \small
 
  \begin{tabular}{llrlp{2.6cm}lp{1.7cm}}
    \toprule
    \textbf{PID} & \textbf{Role} & \textbf{Exp.} & \textbf{Primary Analysis Tools} & \textbf{Analysis Frequency} & \textbf{Coding Comfort} & \textbf{Tasks Shown} \\
    \midrule
    P01 & Data Scientist & 6 Y  & Code/Computational Notebooks  & Daily & Very Comfortable & [T1, T2, T3] \\
    P02 & Software Engineer & 4 Y & Data Visualization Tools & Daily & Very Comfortable  & [T2, T3, T4]\\
    P03 & Consultant &  26 Y & Spreadsheets & Daily & Somewhat Comfortable & [T7]\\
    P04 & Data Analyst & 10 Y & Data Visualization Tools & Weekly & Somewhat Comfortable  & [T5, T6, T7]\\
    P05 & Software Engineer & 10 Y & Code/Computational Notebooks & Weekly & Very Comfortable & [T2, T3, T8]\\
    P06 & Program Manager &  12 Y&  Spreadsheets & Weekly & Very Comfortable & [T4, T7, T9]\\
    P07 & Program Manager &  6 Y& Spreadsheets  & Monthly & Neutral & [T1, T8, T10] \\
    P08 & Software Engineer & 7 Y& Code/Computational Notebook & Weekly & Very Comfortable & [T5, T6, T7]\\
    P09 & Program Manager &  3 Y& Spreadsheets & Daily & Neutral & [T9]\\
    P10 & Tech. Strategist & 15 Y&  Spreadsheets & Monthly & Neutral & [T4, T5, T6, T9] \\
    P11 & Program Manager &  11 Y&  Data Visualization Tools & Daily & Somewhat Comfortable & [T8, T9]\\
    P12 & Program Manager & 5 Y&  Spreadsheets & Daily & Neutral  & [T2, T3, T7] \\
    P13 & Cloud Architect & 5 Y&  Spreadsheets & Daily & Somewhat Comfortable  & [T9, T10] \\
    P14 & Program Manager & 0.5 Y& Code/Computational Notebook & Weekly & Very Comfortable  & [T5, T6]\\
    P15 & Program Manager & 10 Y& Spreadsheets & Daily & Somewhat Comfortable  & [T1, T4]\\
    P16 & Program Manager & 15 Y& Spreadsheets & < Monthly & Never Coded Before & [T7] \\
    P17 & Software Engineer & 15 Y & Code/Computational Notebook & Weekly & Very Comfortable & [T5, T6] \\
    P18 & Program Manager & 7 Y& Data Visualization Tools & Daily & Somewhat Comfortable  & [T9] \\
    P19 & Finance Manager & 13 Y& Spreadsheets & Daily & Neutral  & [T8, T9]\\
    P20 & Program Manager & 3 Y& Data Visualization Tools & Daily & Somewhat Comfortable & [T1, T6, T10]\\
    P21 & Customer Manager & 8 Y& Spreadsheets & Daily & Very Comfortable  & [T4, T9]\\
    P22 & Architect & 5 Y & Code/Computational Notebook & Weekly & Very Comfortable  & [T3, T8, T10]\\
    \bottomrule
  \end{tabular}
  \caption{Participants reported using a variety of tools to perform data analysis in their normal workflows and came from a variety of roles and teams. Over half (13/22) of the participants reported being somewhat comfortable (knowing programming basics and can write simple programs) or less with programming.}
   \label{tab:participants}
\end{table*}

Using our probe and prepared tasks, we conducted a user study to observe analysts' verification workflows of AI-generated analyses. Our goal was not to evaluate our specific interface but rather to understand behaviors that emerge when verification artifacts are available. Three research questions guide our study design and analysis. 
\begin{itemize}
    \item \textbf{RQ1 - Behaviors:} \rqWorkflows
    \item \textbf{RQ2 - Artifacts:} \rqArtifacts
    \item \textbf{RQ3 - Background:} \rqBackground
\end{itemize}

\xhdr{Participants}
We recruited participants from a large data-driven software company and advertised an opportunity to work with AI-generated analyses. Participants were recruited via email based on data analysis interest groups and organization mailing lists. We selected participants with data analysis experience and from diverse backgrounds, including how often they performed data analysis, the types of tools they used for analysis, and their self-reported comfort with programming (Table~\ref{tab:participants}). Participants were compensated with a \$30 USD Amex gift certificate. 

\xhdr{Study Procedure}
We conducted our study remotely with participants sharing their screens over a video conferencing tool. Because our design probe was built as a web application, participants loaded the interface on their computer with their choice of browser. 

The study was structured into three parts lasting approximately 60 minutes total: a tutorial phase ($\sim$10 min), an activity phase ($\sim$40 minutes) in which participants worked through the prepared tasks, and a semi-structured interview phase ($\sim$10 minutes). In the tutorial phase, we presented participants with the following scenario: \textit{Imagine having an intern who is provided with a dataset to address an analysis problem. The enthusiastic intern decides to use an AI assistant to complete the problem, correctly uploading the relevant dataset and crafting a prompt that is their interpretation of the original analysis problem. The AI assistant produces an output but the intern is unsure whether it is correct.}

We clarified that the input to the AI assistant is only the intern's prompt and the uploaded dataset. In particular, we directed analysts not to focus on the quality of the prompt, but on the AI-generated output in response to the prompt. We mention that although the code executes, there might be issues with the information the AI uses or how it interprets the intern's prompt. Next, we presented the participants with the intern's prompt and the assistant's response in our interface (Fig.~\ref{fig:interface}). We asked participants to determine if the AI assistant's answer correctly addressed the initial problem, and if not, point to reasons why they think there was a mistake. To eliminate the challenge of participants having to craft their own prompts—a task that presents its own set of difficulties~\cite{ZamfirescuPereira2023WhyJC}—we chose to assign the responsibility of prompt authoring to the intern. This choice also dissociated participants' personal preferences for how they would write prompts which is not the core focus of our study. 

After introducing the analysis scenario, we gave participants a brief walk-through of the interface (Fig.~\ref{fig:interface}). We then had participants complete a fixed tutorial task to familiarize themselves with the interface (T0 in Table~\ref{tab:tasks}). The task intentionally contained an error in which the AI assistant misinterpreted the values of a column and poorly sorted the answer. This strategy aimed to make participants cognizant of potential assistant errors and to encourage vigilance in subsequent tasks. 

Next, in the activity phase, participants read the analysis goal and reviewed the work of the assistant and intern. We assigned tasks to participants to ensure that, collectively, all tasks were covered (Table~\ref{tab:tasks}). For all tasks in the study, the motivation, analysis goal, and description of the dataset were always available in a left side panel (see appendix for an example). To make sure the participants fully understood the analysis goal, we discussed any areas they found unclear. We also clarified any questions about the dataset or analysis goal while participants were working through the task. To reduce time pressure and encourage a realistic workflow, we emphasized the exploratory nature of our study, allowing participants to work at their own pace until they were satisfied with their conclusion. After completing each task, participants were asked to complete a brief survey that asked for a confirmation as to whether the AI's generation was erroneous 
and their reasoning.



While participants were completing the tasks, we encouraged them to think aloud and describe their reasons for interacting with different components of the interface. If they remained relatively silent, we  regularly prompted them to speak about their workflow. We provided minimal help beyond clarifying the analysis goal or semantics of the data. Finally, in the semi-structured interview, we asked open-ended questions to understand participants' backgrounds and prior experiences, overall verification workflows, and rationales behind their behaviors.\enlargethispage{12pt}



\xhdr{Transcript and Workflow Analysis}
\begin{table}[t]
\small 
    \centering
    \begin{tabular}{lp{5.7cm}}
        \toprule
        Label & Definition \\
        \midrule
         \fboxsep=1.2pt{\colorbox{lgray}{\textcolor{tgray}{Start}}}~& Start of a verification workflow \\
         \vspace{2pt}\fboxsep=1pt{\colorbox{lteal}{\textcolor{tteal}{Data Only}}} ~& Using data artifacts only\\
         \vspace{2pt}\fboxsep=1.2pt{\colorbox{lblue}{\textcolor{tblue}{Data + Procedure}}}~& Mainly focusing on the data and using procedure artifacts as secondary support \\
         \vspace{2pt}\fboxsep=1pt{\colorbox{lred}{\textcolor{tred}{Procedure Only}}}~& Using procedure artifacts only \\
         \vspace{2pt}\fboxsep=1.2pt{\colorbox{lorange}{\textcolor{torange}{Procedure + Data}}}~& Mainly focusing on the procedure and using data artifacts as secondary support \\
         \vspace{2pt}\fboxsep=1.2pt{\colorbox{lyellow}{\textcolor{tyellow}{Notice}}}~& Noticing an issue that may affect the correctness of the AI-generated result \\
         \vspace{2pt}\fboxsep=1.2pt{\colorbox{lgreen}{\textcolor{tgreen}{Confirm}}}~& Confirming or rejecting that the AI-generated analysis has an error\\        
        \bottomrule
    \end{tabular}
    \caption{\textbf{Definition of labels used to annotate participants' verification workflows.}  \fboxsep=1.2pt{\colorbox{lteal}{\textcolor{tteal}{Data Only}}}~and~ \fboxsep=1.5pt{\colorbox{lblue}{\textcolor{tblue}{Data + Procedure}}}~are behaviors under \textit{data-oriented behaviors}~while  \fboxsep=1.2pt{\colorbox{lred}{\textcolor{tred}{Procedure Only}}}~and~ \fboxsep=1.5pt{\colorbox{lorange}{\textcolor{torange}{Procedure + Data}}}~are behaviors within \textit{procedure-oriented behaviors}.}
    \label{tab:labeldefs}
\end{table}
To identify recurring themes, two authors reviewed the participants' screen recordings,  engaged in several rounds of open coding, and had multiple meetings to refine their observations. The themes we observed highlighted variations in participants' workflows and the artifacts they used. In the analysis and results, we distinguish between \textit{data artifacts} (those in Fig.~\ref{fig:interface} right) and \textit{procedure artifacts}~(those in Fig.~\ref{fig:interface} left). Data artifacts pertain to the original, intermediate, and result data (i.e., the data table and summary visualization). Conversely, procedure artifacts encompass the natural language explanation, code/comments, and the AI's explanation and interpretation of its result.

\begin{figure*}[t]
  \centering
  \includegraphics[width=1.0\linewidth]{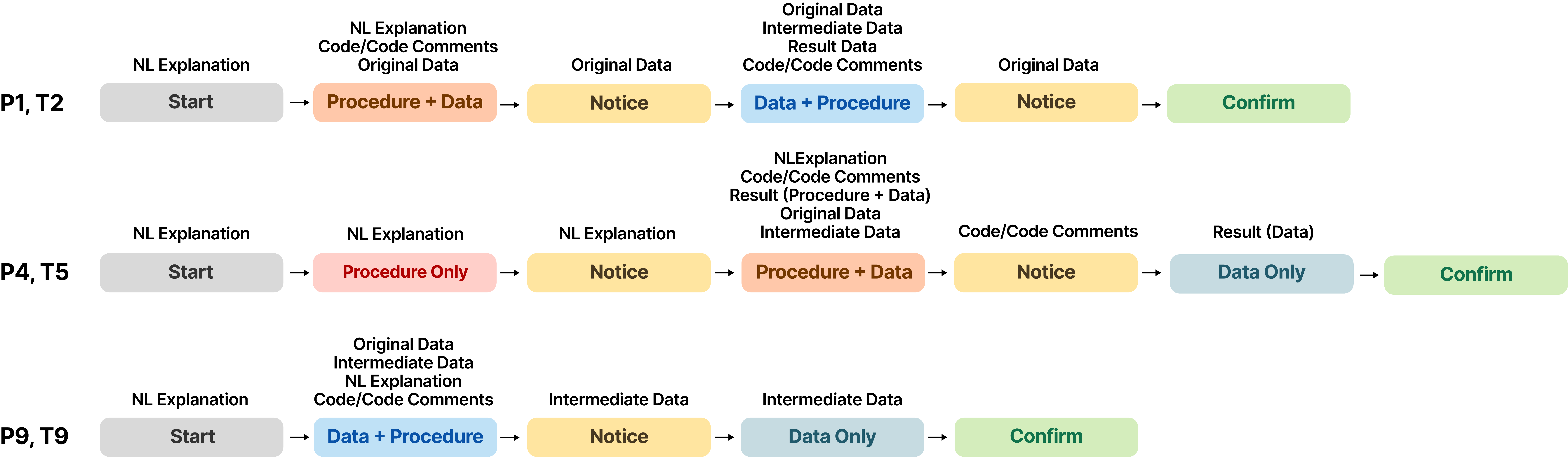}
  \caption{We show examples of interesting end-to-end verification workflows and associated artifacts using our labels in Table~\ref{tab:labeldefs}. The labels help get a sense of the overall workflow and capture relevant behavioral patterns. For example, in T5, we observed P4 starting out (\fboxsep=2.0pt{\colorbox{lgray}{\textcolor{tgray}{Start}}}) focusing only on the natural language explanation (~\fboxsep=1.2pt{\colorbox{lred}{\textcolor{tred}{Procedure Only}}}~) before noticing an issue (~\fboxsep=1.5pt{\colorbox{lyellow}{\textcolor{tyellow}{Notice}}}~) from the explanation. Their behavior then shifts slightly as they include data artifacts (~\fboxsep=2.0pt{\colorbox{lorange}{\textcolor{torange}{Procedure + Data}}}~) before noticing a subsequent issue in the code (~\fboxsep=2.0pt{\colorbox{lyellow}{\textcolor{tyellow}{Notice}}}~). Finally, they check the result data (~~\fboxsep=1.2pt{\colorbox{lteal}{\textcolor{tteal}{Data Only}}}~) to confirm an error in the AI's analysis (~\fboxsep=2.0pt{\colorbox{lgreen}{\textcolor{tgreen}{Confirm}}}~). Overall, we observed 52 verification workflows in our study (39 of which involved errors) with an average length of 4.40 (std=1.42) labels. Four of these had two ~\fboxsep=2.0pt{\colorbox{lyellow}{\textcolor{tyellow}{Notice}}}~patterns occur, 30 had one~\fboxsep=2.0pt{\colorbox{lyellow}{\textcolor{tyellow}{Notice}}}~pattern occur and the rest had none.
  }
  \Description{Three linear flow charts from left to right connected by arrows for a participant and task pair. The first row is for P1, T2, the second row for P4, T5, and the third row for P9, T9. Behaviors are written in blocks connected by arrows. Above the blocks are different artifacts that were used for that behavior. The start states are all 'Start', and the end states are all 'Confirm'.}
  \label{fig:cases}
\vspace*{-3pt}
\end{figure*}

Additionally, to systematically understand participants' verification workflows, we delineated these into sequences of primary observations. 
Informed by the themes from open coding, the two authors who developed the themes drafted an initial set of observation labels (e.g., \textit{read data table} or \textit{raising a consideration}). Both authors annotated a few participants' workflows independently before meeting to resolve ambiguities and unclear interpretations. These discussions facilitated a refinement of the observation labels by broadening their scope. Using the final refined set of labels~(Table~\ref{tab:labeldefs}), they rewatched the recordings and annotated all end-to-end verification workflows in the study (see Fig.~\ref{fig:cases} for examples), including observations of the artifacts being used. 

Each workflow starts when participants began verifying the AI-generated analysis (after understanding the task and data) and ends when they confirmed or rejected the existence of an error. We list all participants' verification workflows in the appendix.  

\section{Results}
\label{sec:results}

\begin{figure}[t]
  \centering
  \includegraphics[width=\linewidth]{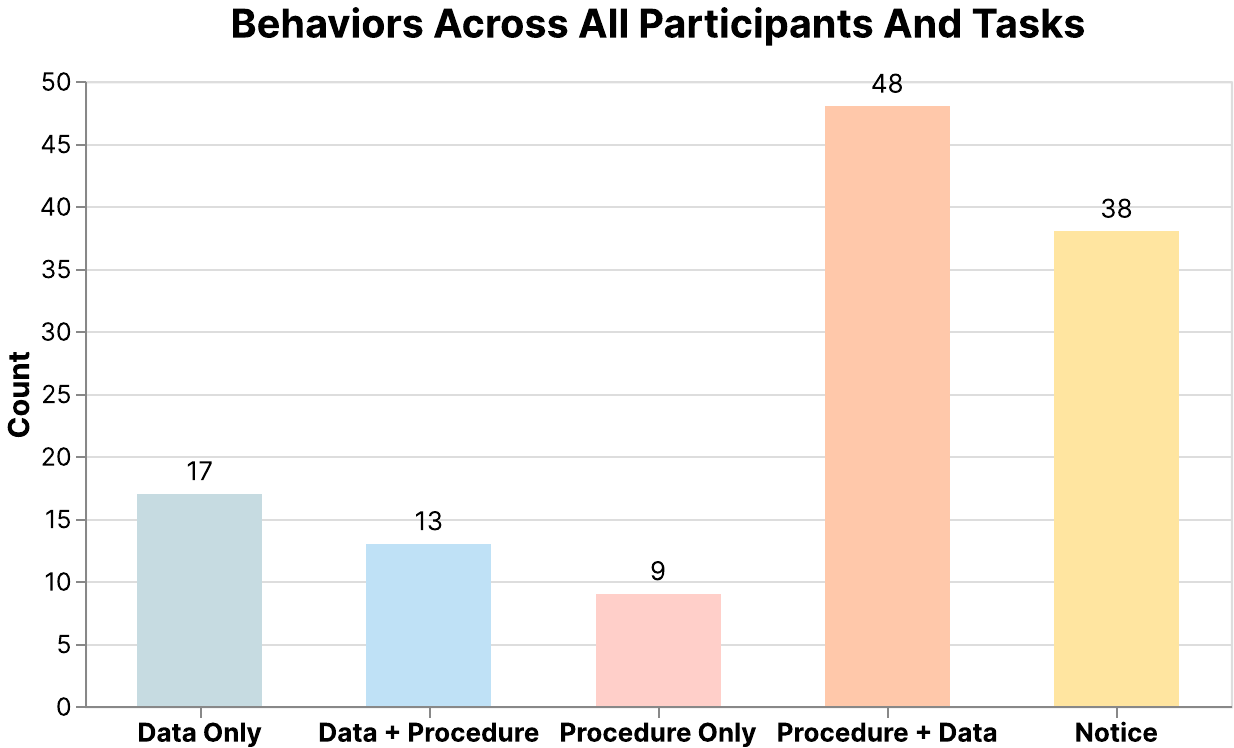}
  \caption{\textbf{Participants often followed \textit{procedure-oriented behaviors}.}}
  \Description{Bar chart of the counts of different behaviors. 'Data Only' has a count of 17, 'Data + Procedure' has a count of 13, 'Procedure Only' has a count of 9, 'Procedure + Data' has a count of 48, and 'Notice' has a count of 38.}
  \label{fig:workflows}
\end{figure}

Participants engaged in a variety of different steps to identify potential issues and ultimately confirm or reject the AI's analysis. These were guided by two high-level \behaviors participants had within a verification workflow:~\textit{procedure-oriented \behaviors} and \textit{data-oriented \behaviors}~(Fig.~\ref{fig:workflows}).\enlargethispage{12pt}

Within a verification workflow, participants with \processworkflows~were primarily focused on validating the AI's procedure, possibly drawing upon data artifacts to aid their understanding. In contrast, participants exhibiting \dataworkflows~gave precedence to scrutinizing the data involved in the analysis, relegating the AI's procedure to a supplementary role. The distinction between these \behaviors lies in the participants' intention. Participants following \processworkflows were concerned with \textit{what does the AI do?} Meanwhile participants following \dataworkflows focused on \textit{does the data make sense?} This was clear from observing their interaction patterns and verbal comments. For example, while some participants engaged extensively with intermediate datasets, their main goal in those moments was to confirm the correctness of the AI's procedure; this behavior was labeled as a \processworkflow with the use of data artifacts as secondary support (i.e., ~\fboxsep=1.5pt{\colorbox{lorange}{\textcolor{torange}{Procedure + Data}}}~).
In this section, we provide a detailed account of findings pertinent to our research questions.

\subsection{\rqWorkflows}
\label{sec:results_workflow}

Overall, we observed most participants (18/22) followed all or parts of the AI's procedure at some point during their verification workflows (44/52). In workflows in which participants did not follow the AI's procedure, they only used the analysis goal, intern's prompt, and data (original and intermediate) artifacts. Likewise, most participants (18/22) had \dataworkflows in their verification workflows (25/52). Participants generally exhibited different behaviors at different moments in their workflows to focus on different parts of the AI-generated response.

\subsubsection{Analysts initially gravitate toward \processworkflows}
Participants usually had \processworkflows~initially during their verification workflows (41/52). These \behaviors were relatively time-consuming and similar to when programmers thoroughly examine the code's logic when they validate suggestions from an AI programming assistant. However, in contrast to the programming assistant setting, \processworkflows were rarely as brief as the quick verification methods (e.g., visual inspection and pattern matching) programmers most commonly employ~\cite{Barke2022GroundedCH, Liang2023ALS}. \enlargethispage{12pt}


\begin{table} 
    \small
\begin{tabular}{llp{4.9cm}}
    \toprule
    Patterns & Count & Occurrences in Verification Workflows \\
    \midrule
   \colorbox{lorange}{\textcolor{torange}{PD}}---\colorbox{lyellow}{\textcolor{tyellow}{N}}---\colorbox{lteal}{\textcolor{tteal}{D}} & 9 & T1 [P1], T2 [P5, P12], T4 [P2], T5 [P4], T6 [P14], T8 [P22], T9 [P6], T10 [P22]\\
    \vspace{1.5pt}\colorbox{lorange}{\textcolor{torange}{PD}}---\colorbox{lyellow}{\textcolor{tyellow}{N}}---\colorbox{lorange}{\textcolor{torange}{PD}} & 6 & T3 [P5, P12], T6 [P4, P14], T7 [P12], T10 [P20] \\
    \vspace{1.5pt}\colorbox{lorange}{\textcolor{torange}{PD}}---\colorbox{lyellow}{\textcolor{tyellow}{N}}---\colorbox{lgreen}{\textcolor{tgreen}{C}}& 5 & T1 [P7], T2 [P2], T4 [P6], T5 [P10, P14] \\
    \vspace{1.5pt}\colorbox{lorange}{\textcolor{torange}{PD}}---\colorbox{lyellow}{\textcolor{tyellow}{N}}---\colorbox{lblue}{\textcolor{tblue}{DP}} & 3 & T2 [P1], T4 [P21], T10 [P13] \\
    \vspace{1.5pt}\colorbox{lred}{\textcolor{tred}{P}}---\colorbox{lyellow}{\textcolor{tyellow}{N}}---\colorbox{lorange}{\textcolor{torange}{PD}} & 3 & T5 [P4, P8], T8 [P11]\\
    \vspace{1.5pt}\colorbox{lblue}{\textcolor{tblue}{DP}}---\colorbox{lyellow}{\textcolor{tyellow}{N}}---\colorbox{lteal}{\textcolor{tteal}{D}} & 3 & T4 [P15], T7 [P16], T9 [P9] \\
    \vspace{1.5pt}\colorbox{lblue}{\textcolor{tblue}{DP}}---\colorbox{lyellow}{\textcolor{tyellow}{N}}---\colorbox{lgreen}{\textcolor{tgreen}{C}} & 2 & T2 [P1], T10 [P13] \\
    \vspace{1.5pt}\colorbox{lred}{\textcolor{tred}{P}}---\colorbox{lyellow}{\textcolor{tyellow}{N}}---\colorbox{lred}{\textcolor{tred}{P}} & 2 & T7 [P4, P6] \\
    \bottomrule
\end{tabular}
  \caption{\textbf{Participants' behaviors around noticing an issue.} We report patterns around noticing an issue that occurred more than once in our study. 
  Participants often followed \processworkflows~before noticing an issue and used a variety of \dataworkflows~and~\processworkflows~afterwards to validate their concerns.}
  \label{tab:freq-of-workflows}
\end{table}

\subsubsection{Analysts shift toward \dataworkflows~ once they noticed an issue}
Upon identifying a concern, participants' strategies diversified, delving deeper to explore and validate the perceived issue~(Table~\ref{tab:freq-of-workflows}). Of the 31 times in which participants noticed an issue but had not yet confirmed an error, 17 times (15/22 participants) they continued their verification workflow with \dataworkflows. In particular, 13 of 17 of these occurrence were when participants switched from \processworkflows to \dataworkflows after noticing an issue.\enlargethispage{12pt}

Some participants (P2, P15, P16, P20, P22) identified the source of the issue through eyeballing different data tables, while others had ~\dataworkflows~to check more closely the potential source of an issue (P4, P5). In T2, P5 correctly noticed that there was no consideration for \textit{Nivea Men}. To explore this further, they focused on the original data table, filtering for rows in which the brand was \textit{Nivea} and applying additional filters to spot check specific products. Similarly, to affirm their suspicions of an error, P4 investigated the final result table after noticing the AI may not have grouped rows by \textit{flight code} (Fig.~\ref{fig:cases} second row) [T5].

\subsubsection{Analysts adopt \processworkflows~to get a high-level understanding of the analysis steps}
\label{sec:results_rq1_procedure_high}

 Many participants looked over the AI's procedure to get a high-level understanding of the analysis steps (P1, P2, P4, P8, P11, P17, P18, P20). For instance, in T9, P18 wanted to understand how the AI arrived at a particular actor-director pair in the result: \shortquote{I'm trying to read the code and I'm trying to understand ... like how is it generating that pair and if there is any error in that process.} Sometimes, participants gained confidence in the AI's correctness
from this high-level understanding  (P1, P2, P11, P20). For example, P1 confirmed their understanding after reading the AI's  procedure one last time:~\shortquote{seems reasonable to me.} P11 was reassured by the AI's overall procedure matching their own:~\shortquote{the steps it described is exactly what I would have done but in SQL.} In other cases, participants' focus on the procedure helped them notice and subsequently confirm a particular step the AI did wrong (P4, P8). For example, P4 and P8 both noticed from the NL explanation that the AI's order of analysis steps were incorrect [T5]. In T6, similar to P11, P4  broke down the general steps on their own and realized the AI missed calculating the average after inspecting the AI's procedure.

\subsubsection{Analysts adopt \processworkflows~to confirm details in the data operations}
\label{sec:results_rq1_procedure_low}
Participants also focused on lower-level details involved in the AI's data operations  (P2, P7, P8, P9, P10, P11, P13, P15, P17, P22). Participants checked functions used in the code, logic around the data operations, and specific result values. P13 was especially vigilant making sure that \shortquote{the spelling was correct} in the data columns mentioned in the code, and that \shortquote{syntactically ... (they) appear to be correct.} P19 was also following the AI's data operations to see how it handled unusual values the AI had identified earlier in its procedure. Focusing on low-level details was one way participants gained confidence in the AI or noticed issues (P7, P8, P10, P11). For instance, in T9, P10 focused on the AI's data operations in the code which helped them conclude the AI had indeed found the highest count. P11, while not the most comfortable with Python, used code comments to translate the AI's procedure into familiar data operations, noticing the AI's procedure was inconsistent with the analysis goal description [T8].

However, participants sometimes formed incorrect conclusionsa due to their reliance on the AI's procedure (P15, P17). In T5, P17 was initially skeptical of the result table due to insufficient information presented, but gained confidence that the result \shortquote{looked more or less correct} after cross referencing the AI's NL explanation and code with the data. P15 manually reviewed the original data and used the AI's NL explanation and code to form an incorrect expectation of the output. They confirmed that the AI was correct when the result data table matched their expectation [T1].

\subsubsection{Analysts adopt \dataworkflows~to better understand the structure of and relationships in the data}
\label{sec:results_rq1_data}
Participants in \dataworkflows~focused on the data in a variety of ways. Some participants (P3, P7, P15, P19) were keen on understanding the data (original and intermediate) and the specific columns that were pertinent to the analysis goal. For instance, P7 adopted \dataworkflows~ using only data tables at the beginning of their process, exploring the column of interest in the analysis goal and subsequently identifying a data quality issue in the frequency column [T10]. Meanwhile, for P9, a prototypical data-oriented participant, their \dataworkflows~involved comparing all data tables [T9]. Specifically, they delved into the directors and actors intermediate data tables to gather an expectation of what they would expect in the subsequent merged data table (Fig.~\ref{fig:cases} third row). After, they checked if specific actors in the result data existed in the merged data. In the process, they spent significant time forming an expectation of the result data, transitioning back and forth between all data tables.

\subsection{\rqArtifacts}
\label{sec:results_artifacts}
\begin{figure*}[t]
  \centering
  \includegraphics[width=0.95\linewidth]{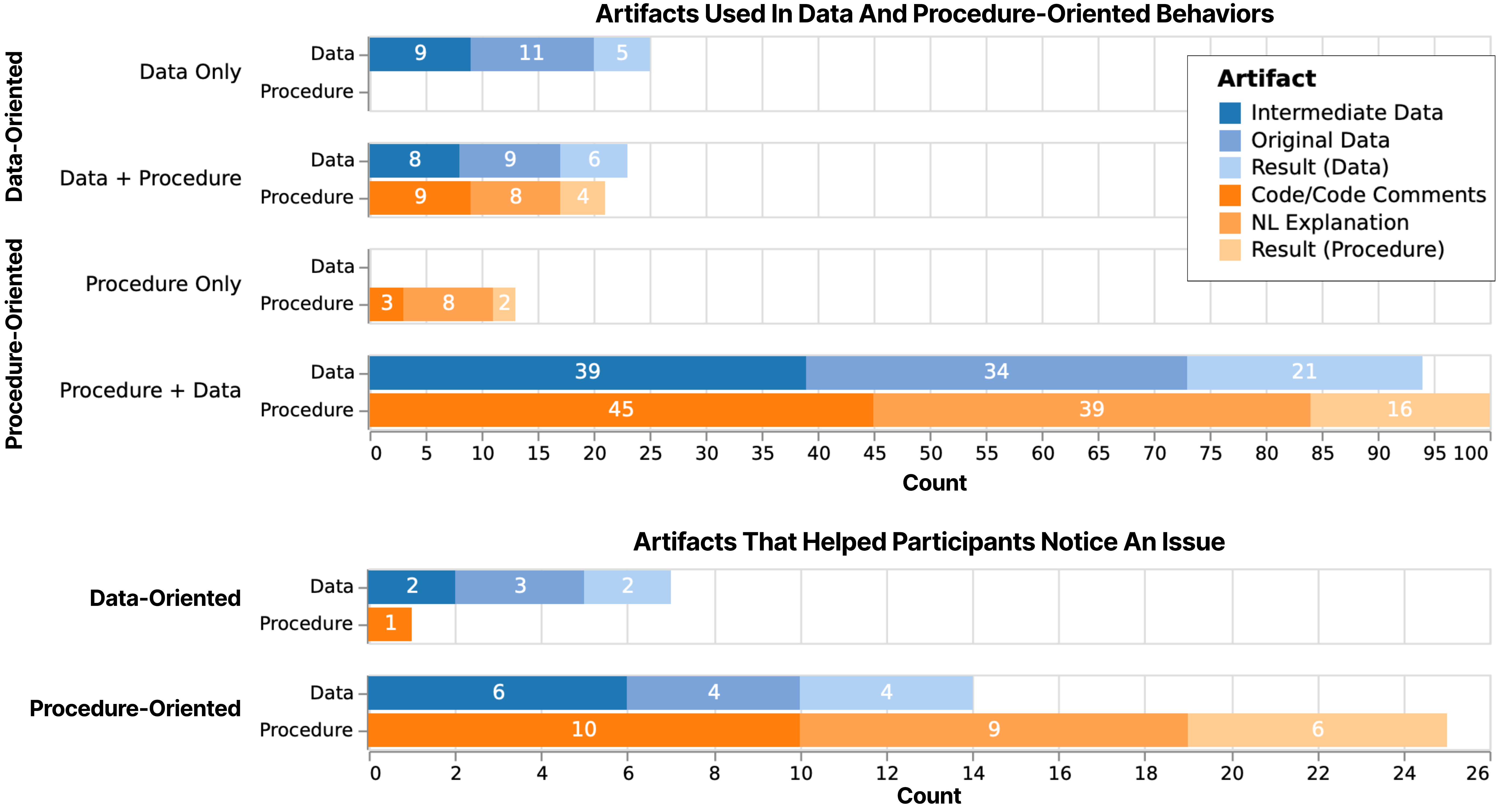}
  \caption{Both data and procedure artifacts were used to support participant's primary behaviors. For each type of behavior illustrated in Fig.~\ref{fig:workflows}, we tally the unique artifacts involved, counting each artifact once per occurrence in a behavior. This distribution shows that participants extensively used data artifacts to support \processworkflows~and used procedure artifacts to support \dataworkflows~ (Top). The intermediate data, original, data, code, and natural language explanation were all pivotal for analysts' to notice an issue (Bottom).}
  \Description{Two vertically stacked horizontal bar charts of counts of artifacts used in data and procedure-oriented workflows. The top one shows the total counts used in the data and procedure-oriented behaviors, and the bottom one shows the counts used in different behaviors that helped participants notice an issue.}
  \label{fig:artifacts}
\end{figure*}

Across all participants, both data and procedure artifacts were used consistently in their verification workflow (Fig.~\ref{fig:artifacts}). Data artifacts were frequently employed to support \processworkflows~(17/22 participants and in 41/52 workflows), while procedure artifacts were utilized during \dataworkflows (10/22 participants and in 13/52 workflows).  Within the data, participants engaged with the intermediate data roughly equally as often as the original data and result data. Similarly, for procedure artifacts, participants leveraged code and code comments as often as the natural language explanation. However, these preferences for artifacts were not consistent between participants (see Sec.~\ref{sec:results_bg}).

\subsubsection{Analysts use data artifacts for sensemaking}
\label{sec:results_rq2_data}
Unsurprisingly, most participants leveraged data artifacts to understand and interpret the data, a standard practice in data analysis practices~\cite{Alspaugh2019FutzingAM, Batch2018TheIV, Grolemund2014ACI}. For instance, P7 used the data summary visualizations to understand the data, and expressed liking \shortquote{the summary stats to be able to do a quick view of the data and understand what it is instead of having to manually check everything.} P3, meanwhile,  meticulously examined the original, intermediate, and result data to have a comprehensive understanding of the analysis. Consequently, several participants expressed appreciation for the data table summary and visualizations provided in the AI's procedure (P5, P7, P13, P17, P20).  

Through closely examining the data tables, participants were able to identify issues with the AI-generated analyses (P1, P3, P7, P9, P13, P15, P16). For instance, from the data tables, P1 noticed that there was also the \textit{Nivea Men} brand in the dataset which the AI missed (Fig.~\ref{fig:cases} first row), ultimately leading them to confirm that the AI-generated analysis contained an error. Likewise, for P9, their exploration of the data tables led them to perceive a discrepancy with the number of rows in merged data table and the count from a quick mental calculation. These observations align with established guidelines in data analysis practices and implementations of data analysis tools, and underscore the importance of data visualizations alongside the analysis procedure~\cite{Alspaugh2019FutzingAM, Kery2019TowardsEF, Choi2023TowardsTR, Batch2018TheIV, Kery2018InteractionsFU, Wu2020B2BC, Moritz2017TrustBV} .

\subsubsection{Analysts frequently use data artifacts for both quick checks and detailed calculations to support \processworkflows}
\label{sec:results_rq2_data_support_procedure}

Some participants glanced at the data artifacts to quickly check things they noticed in the AI's procedure (P6, P8, P22). For instance, P22 noticed that the AI's calculation for the average price of gourmet food was not a number, prompting them to check the original data table and discover the AI used the wrong column name [T2]. Similarly, participants used data artifacts to conduct data sanity checks to verify the procedure (P1, P3, P4, P5, P19, P20, P21, P22). For instance, P19 wanted to see if percentages in a column added up to 100: \shortquote{If the percentages tied to 100\%, I would feel confident just at a glance that you know everything and action was allocated to one of those languages} [T8]. P20, meanwhile, went back and forth between the code and table to verify the procedure, stating, \shortquote{just like understanding what the code did and then each of the intermediate steps in seeing like ... it looks like it joined it} [T6].
Alternatively, participants also tried to replicate the AI's data operations using the supported functionalities in the data tables to compare with the result in the AI's procedure and verify what the AI did (P2, P9, P12, P14, P22). For example, P22 filtered on the \textit{frequency} column following the AI's steps to see how many rows were filtered out [T10]. 

These approaches align with common AI-assisted programming verification methods~\cite{Liang2023ALS, Barke2022GroundedCH}, resembling visual inspection and pattern matching for quick data checks, and code execution for replicating the AI's data operations. However, distinct from AI-assisted programming, data (in addition the procedure) serves as a central artifact for verification.\enlargethispage{12pt}

\subsubsection{Analysts use procedure artifacts to find discrepancies in the procedure}
\label{sec:results_rq2_procedure}

Some participants were keen on verifying the consistency between the high-level analysis steps in the NL explanation and the specific data operations in the code (P5, P11, P17, P22). Meanwhile, other participants compared the NL explanation with their own common knowledge. For example, P10 felt that \shortquote{2800 flights felt like a lot} for a day [T5], leading them to investigate the AI's procedure. Similarly, from inspecting the AI's explanation of the result, P6 noticed \shortquote{Jurassic World is not a Bollywood movie} [T4]. These discrepancies prompted them to question the AI's data operations and understanding of the dataset. Moreover, participants used procedure artifacts to verify the AI's interpretation of the prompt and original dataset (P4, P12). As a result, in T7, both P4 and P12 noticed issues with the AI's understanding of the data, stating an incorrect interpretation of a column.\enlargethispage{12pt} 

\subsubsection{Analysts use procedure artifacts for tracing data provenance to support \dataworkflows}
\label{sec:results_rq2_procedure_support_data}
Many participants (P1, P3, P13, P15, P16) used the NL explanation and code to trace the provenance of data. P1 in T2 is a typical example (Fig. \ref{fig:cases} first row). P1 started out with~\processworkflows~but upon perceiving an issue in the original data table where more than 5 items had a rating of 5, transitioned into \dataworkflows. They scrutinized the original data table but became confused when the result data table displayed values less than 5. This contradicted their initial observation where more than five products had a 5 rating in the original data. To resolve this, they turned to the intermediate data and the AI's procedure, tracing the steps in the code that derived the result table. 

Participants also noticed issues when using procedure artifacts to support their understanding of the provenance of the data (P1, P9, P13, P15, P16, P17, P21). For instance, P17 wanted to explore why the intermediate data table contained all kinds of departure time. They checked the code to see where it did the filtering on the \textit{Departure Time} and realized the AI missed taking into account the time of the day when calculating the minimum of prices [T6]. These behaviors reiterate the importance of tracking data provenance in analysis tools and highlight opportunities to adopt and adapt methods in existing implementations~\cite{Kery2017VarioliteSE, Rule2018AidingCR, Kery2018TheSI, Wang2022DiffIT, Pu2021DatamationsAE, Wu2020B2BC, Park2021StoryFacetsAD}.

\subsection{\rqBackground}
\label{sec:results_bg}
In the post-task interviews, we explored participants' prior experiences with data analysis, coding, and existing analysis tools. We noted a range of experiences that might be associated with participants' behaviors and interactions with artifacts.

\subsubsection{Prior Experience With Data}
\label{sec:results_bg_data}

Some participants with \dataworkflows~ mentioned their prior experience working with data (P6, P9, P11, P16, P17, P18, P22). For instance, P15, who had previous experience working with large datasets, expressed comfort with directly manipulating the datasets, stating, \shortquote{I'm used to working with hundreds of thousands of rows, so the size of the spreadsheet wouldn't have been a problem.} In both tasks P15 did, they only followed \dataworkflows. 

These prior analysis experiences also led participants to be vigilant about the data (P6, P9, P11, P22). P6, who self-identified as a \shortquote{data guy}, was inclined to dive directly into the data and scrutinize the tables and numbers. P11, an electrical engineer by training, also favored a data-oriented approach. 
 They compared the study tasks to their past experiences designing systems, explaining how they always \shortquote{went straight to the data} and asked \shortquote{hey, can I trust that it's quality data?}
 P9, with a background in finance and accounting, was accustomed to handling datasets that were \shortquote{routinely between 100 and 500,000 rows.} In the study, P9 focused on validating the data, underscoring the importance of ensuring data integrity and rational calculations. Similarly, P22  cited their  prior work experiences as informing their analysis verification approach: \shortquote{we were maniacal about not drawing the wrong conclusion.} During their verification workflows, they demonstrated a methodical approach to problem-solving, routinely considering \shortquote{lots of caveats and checks.}  

Moreover, we noticed participants' prior experiences forming conclusions and insights from data shaped their behavior in the study tasks. Participants  (P1, P6, P9, P19, P22) who started their verification workflow by proactively devising a plan mentioned being accustomed to formulating error hypotheses, setting expectations, and actively identifying problems in the data. For example, P19 commented how in their own data analysis workflow they were always on the lookout for \shortquote{weird stuffs} like data classes, data types, and anomalies. This was reflected in their systematic approach when faced with our study tasks.\enlargethispage{12pt}

\subsubsection{Prior Experience With Code}
\label{sec:results_bg_code}
While participants' experiences working with data sometimes led them to being more data-oriented, a couple participants' coding comfort made them inclined towards \processworkflows and examining the code (P5, P8). For instance, P8, a software engineer, naturally tried to verify the correctness of the code: \shortquote{as a developer, just let me look at the code.} Similarly, P5, compared the tasks to doing a \shortquote{code review.}
Interestingly, participants who were not familiar or comfortable with Python but had a familiarity of data operations and/or data-oriented programming languages (e.g., SQL and Visual Basics for Applications) managed to leverage their knowledge to understand the AI's procedure on the data (P10, P15, P19, P20, P21). For example, P4 looked for key words in the code to understand the the AI's operations. For P10 in T9, despite having no prior experience with Python, they were able to devise the analysis steps in SQL and cross-check them against the AI's explanation and code comments: \longquote{If I were to do this in SQL Server with the table that you gave me, I would have created ... select unique ID where director and popped the table out ... So basically, I'm translating Python into the technologies I know to see if the AI is doing what I need them to do.} 

\looseness-1 However, while able to parse key words or main points from the code, a lack of understanding of the code hindered some participants' confidence (P2, P12, P14, P18, P20, P21). For example, P20 found T5 \shortquote{difficult because (they didn't) fully understand the code.}
For P18, an incomplete comprehension in T9 led them to focus more on the original data as opposed to the intermediate data since they were unsure how the intermediate data was generated: \shortquote{in the intermediate steps, I am not confident if the slicing and the dicing is being done the way it should be.} To bridge this gap, a couple participants mentioned liking code documentation or being able to easily translate the code to a more familiar languages (P11, P21). For P12 and P21 in particular, a lack of code comprehension led them to prefer the NL explanations over the code. P12 found the explanations to be \shortquote{a very concrete and precise way that makes sense.}

\subsubsection{Prior Experience With Existing Analysis Tools}
\label{sec:results_bg_tools}

Our study aimed to include participants with a diverse usage of existing analysis tools~(Table~\ref{tab:participants}).  While there were ways to verify the AI-generated output using our design probe~(Fig.~\ref{fig:interface}), some participants expressed a preference for functionality they were familiar with from existing tools that were unsupported by our interface (P12, P13, P15, P18, P21). For example, P12 mentioned wanting to use conditional sorting on the columns, and P17 wanted support for  comparison operators in data filtering. Several participants also expressed a desire to have pivot tables (P11, P17, P18, P21), making it easier to understand the data. Moreover, while many participants expressed appreciation for the visual summaries of the data tables (P5, P7, P13, P17, P20), some participants (P4, P18, P21) expressed low trust in the data artifacts because they were unsure how the artifacts were created in the first place. P21 in particular stated: \longquote{The (visualization) summary tab was kind of useless to me... I didn't really trust it and and it was unclear to me how the the data was being summarized.}\enlargethispage{12pt}



\section{Discussion}
\label{sec:discussion}
In this work, we examined how data analysts understand and verify AI-generated analyses. We developed a design probe, prepared realistic analysis queries and AI-generated analyses, and conducted a user study observing analysts' verification workflows and common behaviors. In this section, we synthesize the results from our user study and share implications for data analysts using AI-based analysis tools (Sec.~\ref{sec:discussion_impl_analyst}) and system designers building these tools (Sec.~\ref{sec:discussion_impl_designers}).

\subsection{Implications for Data Analysts}
\label{sec:discussion_impl_analyst}
\subsubsection{The Role of Data Literacy in Analysis Verification}
\label{sec:discussion_impl_analyst_data_lit}
Our findings suggest that while a lack of familiarity with one particular programming language (i.e., Python) is not a barrier for verifying AI-generated analyses, prior experience with data operations in other languages (e.g., SQL or Visual Basics for Applications) can aid in understanding the AI's procedure (Sec.~\ref{sec:results_bg_code}). Likewise, knowledge and experience in data analysis are often indispensable for grasping AI-generated analyses, mirroring the requirement for computational thinking skills when working with AI-based code assistants~\cite{Sarkar2022WhatII}. Here, we highlight two recommendations for analysts to be more effective in verifying AI-generated analyses.


\yhdr{Know Common Data Operations} We observed participants closely examined the AI's data operations and sometimes performed these operations on the datasets directly to compare with AI's result (Secs.~\ref{sec:results_rq1_procedure_low} and~\ref{sec:results_rq2_data_support_procedure}). Their actions involved understanding common operations such as filtering, sorting, and merging the data. Similarly, understanding aggregation operations, such as sum, average, count, were essential to verify how the AI grouped data and performed calculations on those groups.


\yhdr{Develop Strategies for Quick Data Validations} Our observations revealed analysts often employed quick sanity checks on the data to assess AI-generated outputs (Secs.~\ref{sec:results_rq2_data} and ~\ref{sec:results_rq2_data_support_procedure}). Such strategies involved eyeballing values, manual summation of rows, using visualizations to check the value distributions, spot checking result values against original data table, and comparing things against common knowledge. These were effective for analysts to either build confidence in the results or identify unexpected patterns. The sanity checks often involved forming expectations around the result and then comparing them with the AI-generated output. Any discrepancies signaled potential issues that demanded further inspection. Given the effectiveness, analysts should actively identify quick sanity checks and externalize their expectations~\cite{Russell1993TheCS}. However, as there may be data issues hidden by seemingly reasonable patterns~\cite{correll2018looks}, analysts should employ a variety of different sanity checks.\enlargethispage{12pt}

\subsubsection{The Role of AI Assistants in Data Analysis}
\label{sec:discussion_impl_analyst_ai}

Participants' varied experiences working with data and analysis tools (Sec.~\ref{sec:results_bg}) led to a wide range of comments about their preferred use of general AI-based analysis assistants. Here, we discuss how AI assistants could be used in data analysis.

Some participants commented how AI-assistance would be helpful for getting them a partial answer that might guide them in their analysis (P1, P6, P7, P14, P20, P21). For instance, P6 mentioned how the AI can help them consider alternative steps: \shortquote{It'll help me say `oh maybe I didn't think of this way of solving this and I can now go down that path.'} AI assistants can help inspire ideas for analysis steps similar to when programmers follow an exploration mode in AI-assisted programming~\cite{Barke2022GroundedCH}. Presenting alternative approaches is especially critical in analysis assistance as analysts can often be limited in their consideration of analysis approaches~\cite{Liu2019PathsEP, Liu2020UnderstandingTR}, thereby affecting the robustness of their subsequent analysis conclusions~\cite{Silberzahn2018ManyAO, Breznau2022ObservingMR, Schweinsberg2021SameDD}. Thus, AI analysis assistance can improve analysis quality by broadening the analysis decision space analysts consider ~\cite{Liu2019PathsEP, Gu2022UnderstandingAS}. 

Similarly, participants noted that assistants can expedite their analysis workflow (P1, P13, P20), corroborating the findings from prior research on AI-code assistants ~\cite{Vaithilingam2022ExpectationVE, Ziegler2022ProductivityAO, Dibia2022AligningOM, GitHubCopilotResearch}. However, to ensure the robustness of analysis conclusions, it is crucial to balance the acceleration of analysis with measures that surface potential errors, underlying data assumptions, and alternative approaches~\cite{Gu2023HowDD}.

Finally, some participants expressed a preference to reserve AI assistance for peripheral tasks such as error-checking or report generation rather than the central analysis. Utilizing AI assistance for crafting reports~\cite{Zheng2022TellingSF, Wang2023Slide4NCP} and similar subsidiary activities offers a low stakes environment for analysts to evaluate the quality of AI-generated outputs within data analysis workflows. Such use cases can also facilitate a deeper understanding of the assistant's capabilities and limitations.

\subsubsection{Understanding the Bounds of the AI and the Tool}

Participants sometimes expressed doubt on the intermediate data artifacts as it was unclear what artifacts had the potential to be a result of AI ``hallucinations"~\cite{Ji2022SurveyOH} and which were consistent with the data and procedure (Sec.~\ref{sec:results_bg_tools}). As the boundary blurs between AI-generated artifacts and tool-generated artifacts, analysts should clearly distinguish what is and what is not produced by AI.

Tool artifacts compiled from code is deterministic (i.e., the same summary visualization will be produced given the same code and input data)~\cite{Sarkar2022WhatII}. The AI-generated outputs and their associated errors are stochastic in nature, varying each time the AI is prompted to conduct the analysis. Knowing the provenance of artifacts used to support verification---what is tool-generated (e.g., the visual summary of data) and AI-generated (e.g., the intermediate data feeding into the visualization)---is important to assess the reliability of the information associated with these artifacts. Additionally, understanding the strengths and limitations of the AI~\cite{Ragavan2022GridBookNL, Sun2022InvestigatingEO} is crucial to knowing what types of mistakes to look for.\enlargethispage{12pt}

\subsection{Implications for System Designers}
Informed by our study findings of verification workflows and noted challenges, we highlight four design implications for designers of AI-based analysis assistants. 

\label{sec:discussion_impl_designers}
\subsubsection{Fluidly connecting data-oriented and procedure-oriented artifacts at various levels of granularity}
Our study on analysts' verification workflows indicated the use of data-oriented and procedure-oriented artifacts was often intimately tied to each other, requiring analysts to switch between the two (e.g., intermediate data table and the code snippet that generated this data). Therefore, tools should facilitate a seamless connection between the data and procedure. Integrating the data with the procedure aligns with existing implementations in data science and machine learning tools~\cite{Rahman2020LeamAI, Choi2023TowardsTR, Bauerle2022SymphonyCI, Kery2020mageFM, Wu2020B2BC}. Likewise, as analysts prefer different functionality based on their prior experiences (Sec.~\ref{sec:results_bg_tools}), data components should be interactive, reusable, and customizable to cater to specific preferences and applications~\cite{Bauerle2022SymphonyCI, Rahman2020LeamAI, Choi2023TowardsTR}.

These connections should also prioritize verification behaviors. Participants often explored AI-generated analyses at different levels of abstraction (Secs. ~\ref{sec:results_rq1_procedure_high} and ~\ref{sec:results_rq1_procedure_low}) and transitioned between artifacts for quick glances of specific information (e.g., column names) or in-depth examinations involving multiple steps (Secs.~\ref{sec:results_rq2_data_support_procedure} and \ref{sec:results_rq2_procedure_support_data}). Therefore, tools should allow analysts to specify the scope of their current verification actions and surface the most relevant information. For example, when an analyst selects the text of a column name or cell value in the procedure, the tool can quickly highlight the corresponding data (and vice-versa). If the analyst selects a code snippet containing multiple data operations, the tool can highlight the corresponding data differences between the operations~\cite{Wang2022DiffIT}. Furthermore, to support analysts in tracing the data provenance of a given intermediate data (Sec.~\ref{sec:results_rq2_procedure_support_data}), the tool can highlight just the data operations in the procedure that produced it~\cite{Kery2017VarioliteSE, Head2019ManagingMI}. To support these features on the back-end, the AI can dynamically augment its text response to include semantically relevant connections to the data. Recent systems have demonstrated the efficacy of augmenting the AI's output for AI-assisted writing and exploration~\cite{Jiang2023GraphologueEL, Laban2023BeyondTC}.

\subsubsection{Communicating the semantics of data operations}
\label{sec:discussion_impl_designers_semantics}
We observed AI-generated code to be a common part of analysts' verification workflows in procedure-oriented and data-oriented behaviors~(Secs.~\ref{sec:results_rq2_procedure} and ~\ref{sec:results_rq2_procedure_support_data}). However, a lack of comprehension of the code and the exact data operations impacted analysts' confidence in their verification findings and even pushed them to alternative behaviors~(Sec.~\ref{sec:results_bg_code}). Nevertheless, participants who were able to translate the AI's code into familiar data operations were able to understand the procedure and not be significantly impacted. Therefore, consistent with implications from prior work~\cite{Kim2022HelpMH, Gu2023HowDD}, tools should provide multiple methods to communicate the operations in the AI's procedure.

For instance, participants suggested translating the Python code into programming languages they were comfortable with. This translation can be achieved with neural code translators~\cite{Lachaux2020UnsupervisedTO} or existing AI-code assistants~\cite{ChatGPT, githubcopilot} at the code snippet level. 
If translated at the operation level, it is essential to ensure faithfulness to the original operator. Alternatively, the code can be translated into systematic and consistent natural language utterances representing the data operations. On a limited set of operations and data problems, Liu et al.~\cite{Liu2023WhatIW}  found this approach improved users' understanding of the code-generating model. 
Additionally, besides code and natural language, tools can communicate data operations through visualizations such as those in \textit{Datamations}~\cite{Pu2021DatamationsAE}, and \textit{SOMNUS}~\cite{Xiong2022VisualizingTS}. Likewise, tools can show differences in intermediate data between calculations to further enhance analysts' understanding of the operations~\cite{Wang2022DiffIT}.

\subsubsection{Communicating AI's assumptions and interpretations of the data}

We observed participants often noticed errors faster in tasks where the AI clearly stated its assumptions about the data (e.g., the data column semantics, assumed data types, and relationships between data columns etc.) and its interpretation of the prompt (P6, P10, P12, P22). For example, in T7, the AI clearly stated its interpretation of what the data in key columns represented: \shortquote{2021: (Assuming) Rank of the hotel in 2021} and \shortquote{Past\_rank: Past rank (unclear for which year).} However, we noticed similar assumptions were not always stated in other tasks.\enlargethispage{12pt}

Therefore, tools should consistently communicate the assumptions and interpretations of the dataset and analysis, a well-regarded best practice for data analysis~\cite{Field2012DiscoveringSU}. This implication narrows down a broader recommendation for AI-based code assistants to articulate the generated code within the context of a programmer's specific task and environment~\cite{Sarkar2022WhatII}. One strategy to facilitate regular communication could be to prompt the AI system to enumerate its assumptions explicitly. Another avenue, as exemplified by LLMs such as GPT-4 being good evaluators for tasks like natural language summarization and visualization~\cite{Dibia2023LIDAAT, Lin2022TeachingMT, Liu2023GEvalNE}, can involve employing a separate agent to evaluate the assistant-generated analysis, thereby listing any assumptions, interpretations, or discrepancies.

\subsubsection{Incorporating AI guidance into verification workflows}
In a world where AI can generate data analyses, analysis verification becomes an increasingly important skill~\cite{Tu2023WhatSD}. In our study, we noticed variations in participants' verification approaches, with more experienced participants proactively formulating systematic plans (Sec.~\ref{sec:results_bg_data}). As LLMs show promise for making data analysis approachable to broader audiences, there is an opportunity for tools to guide analysts in their verification workflows. 

For example, to help with quick data checks (Sec.~\ref{sec:results_rq2_data_support_procedure}), LLMs can generate test cases. For instance, if a field represents age, then only values between 0 and around 120 might be appropriate. Furthermore, identifying missing values (which might be 0) might help in the analysis process. However, different than prior work where generated test cases were for general programming problems ~\cite{Tufano2020UnitTC, Chen2022CodeTCG}, test cases for data analyses need to consider the semantics and domain of the data. Specifically, as outputs from data analysis are meant for human understanding and involve complex data structures
(e.g., data tables, plots), there can be alternative ways to present the answer (e.g., different views of the same data table)~\cite{Yin2022NaturalLT, Lai2022DS1000AN, Huang2022ExecutionbasedEF}. Likewise, since AI-generated analyses can involve sequences of data operations, it is pertinent to create test cases for both the final output and intermediate steps.

Additionally, as a key part of verification involves checking for inconsistencies (Sec.~\ref{sec:results_rq2_procedure}), our findings suggest opportunities for AI self-reflection~\cite{Jiang2023SelfEvolveAC, Shinn2023ReflexionAA, Park2023GenerativeAI, Sumers2023CognitiveAF} to help analysts identify potential discrepancies, formulate specific checklists, and improve their data literacy (Sec.~\ref{sec:discussion_impl_analyst_data_lit}). This reflection should be contextualized with external knowledge of existing best analysis practices and the contents of the working analysis. In incorporating AI self-reflection, the interface can visually highlight specific sections of the analysis that require verification. Such visual warnings can help reduce analysts' overconfidence in the AI's procedures~\cite{Laban2023BeyondTC}. However, given the failure modes of LLMs~\cite{Bender2021OnTD, Ji2022SurveyOH}, it is crucial to communicate to the analyst the parts of guidance that are AI generated.

\subsection{Limitations and Future Work}
\label{sec:limitation}
We note several limitations in our study and discuss opportunities for future work. \enlargethispage{12pt}

First, the analysis queries involved in the study tasks primarily involved sequences of data transformations. Other aspects of data analysis (e.g., data visualization, statistical modeling, machine learning, etc.) were not explored. However, our study still provided insight into common analysis workflows, as participants (P9, P22) mentioned the issues they encountered in the study tasks were ones they commonly faced in their own data analyses. Our study is a first step towards uncovering verification workflows of AI-generated outputs where data and manipulations on data are centrally involved. Future work should explore analysts' verification of AI-generated data visualizations, statistical analyses, and machine learning pipelines~\cite{Crisan2020PassingTD}.

Second, since our study contained a small number of verification tasks while using real responses generated by state-of-the-art nondeterministic models, components of the AI's explanation (e.g., listing its assumptions of the data and columns) and specific data operations involved in the data analysis task (i.e., filtering for data or merging data) were not controlled for across tasks. Future work could experiment with how varying components of the AI's explanation or analysis task impact analysts' verification processes.

Third, as discussed in our study desisgn considerations (Sec.~\ref{sec:method_design}) participants were not the ones gathering the data and coming up with the analysis goal. As a result, prior to them verifying the AI-generated analyses, they first spent time understanding the data and analysis goal, a process that may not be as pertinent in conducting their own data analysis with AI assistants. However, all tasks used in the study involved real-world datasets with analysis prompts written by data scientists (Sec.~\ref{sec:study}). In addition, we quickly clarified any questions regarding the analysis goal to facilitate analysts' verification. Future work could explore verification workflows when analysts work with an AI assistant on their own data and analysis.

Finally, study participants were not responsible for creating the AI prompts (Sec.~\ref{sec:method_design}). While more elaborate prompts may be argued for reducing the errors presented in our study, writing these prompts is challenging~\cite{ZamfirescuPereira2023WhyJC, Mishra2022HELPMT}. Brevity of the prompt can be in tension with a precise one that reflects all aspects of the analysis~\cite{Yin2022NaturalLT}. 
Therefore, given that the misalignment between the analyst's intent and AI's actions is inherent in natural language communication~\cite{Liu2023WhatIW, Luger2016LikeHA, zcan2020StateOT}, subtle AI-generated analysis errors (at least those due to misalignment) and the need for verification workflows persist. 
In a few cases, our participants (P6, P10, P22) noticed issues directly from language ambiguities in the intern's prompt. As a logical next step, future research should study how expertise in prompt writing influence analysts' interactions with AI assistants and how these analysts integrate prompts into their verification and repair strategies. \vspace{-8pt}  \enlargethispage{12pt}

\section{Conclusion}
In this paper, we explore data analysts' behaviors when verifying AI-generated analyses. We build a design probe and prepare tasks with realistic analysis intent and LLM outputs generated by OpenAI's Code Interpreter. In a controlled user study, we find analysts frequently interleaved \textit{procedure-oriented} and \textit{data-oriented behaviors}, using data artifacts in support of \processworkflows and procedure artifacts in support of their \dataworkflows. We also find analysts' prior analysis and programming experiences influenced their approach and understanding of the AI's output. Based on our findings, we synthesize recommendations for data analysts and highlight opportunities for future tools to enhance the verification experience, including communicating the AI's assumptions and interpretations of the data, and ways to incorporate AI guidance in verification workflows. \vspace{-8pt}

\begin{acks}
We thank analysts from Microsoft for their eager participation in our study and anonymous reviewers for their insightful comments. We also thank Bongshin Lee, Dan Marshall, Dave Brown, Robert DeLine, Priyan Vaithilingam and others from Microsoft Research for their valuable feedback throughout the project. \enlargethispage{12pt}
\end{acks}


\bibliographystyle{ACM-Reference-Format}
\bibliography{0-references_custom}

\appendix

\section{Task Selection Process}

To avoid datasets overlapping with the training data of LLMs~\cite{OpenAI2023GPT4TR, Brown2020LanguageMA}, we chose natural language queries associated with datasets uploaded to Kaggle after February 2022. This included 660 queries from 70 distinct datasets.

We selected a subset of analysis queries that were incorrectly answered by current state-of-the-art LLMs (GPT-4 and Code Interpreter). First, we tested the 660 queries against GPT-4~\cite{OpenAI2023GPT4TR} using its API (temperature=0.7, n=3). For this, GPT-4 was provided a dataset summary similar to that of other LLM-based analysis tools~\cite{Dibia2023LIDAAT}, a scaffold to write code in, and the natural language query. Next, we sampled a set of 50 queries that resulted in incorrect outputs from GPT-4, specifically selecting more complex queries with a high number of Pandas API calls in the solution. These queries (and corresponding datasets) were then presented to Code Interpreter. Code Interpreter executed the code for the analysis, and detailed its steps in natural language. 

After manually inspecting the 50 queries GPT-4 struggled with, we identified 22 that were also incorrectly answered by Code Interpreter. These errors included bad calculations, semantic misinterpretations of the prompt, and misunderstandings of the data. We then chose a subset of these 22 queries that cover a variety of error types (i.e., misunderstanding column semantics, bad calculations, or misunderstanding the query) and involve datasets and domains that were approachable to general audiences. Using this subset, we constructed the tutorial and primary tasks of our study. Sometimes, Code Interpreter encountered ambiguities and sought clarification from the user. For these cases, we included a follow-up prompt and integrated the multi-turn interaction into our study tasks.\enlargethispage{24pt}

\subsection{Example of Dataset Summary used for GPT-4 Prompt}
The following is an example of the dataset summary used as part of the prompt for GPT-4 for the Amazon Orders dataset (T1),

{\footnotesize\begin{lstlisting}[language=json, caption={}, label={lst:json}]
{
  "name": "",
  "file_name": "",
  "dataset_description": "",
  "fields": [
    {
      "column": "order_no",
      "properties": {
        "dtype": "string",
        "samples": [
          "171-5463316-4433940",
          "407-6814126-3628337"
        ],
        "num_unique_values": 171,
        "semantic_type": "",
        "description": ""
      }
    },
    {
      "column": "order_date",
      "properties": {
        "dtype": "date",
        "min": "2021-06-13T19:08:00",
        "max": "2022-02-25T20:44:00",
        "samples": [
          "2022-01-27T17:31:00",
          "2021-10-15T20:27:00"
        ],
        "num_unique_values": 171,
        "semantic_type": "",
        "description": ""
      }
    },
    ...
    {
      "column": "sku",
      "properties": {
        "dtype": "category",
        "samples": [
          "SKU:CR-6E69-UXFW",
          "SKU:ST-27BR-VEMQ"
        ],
        "num_unique_values": 54,
        "semantic_type": "",
        "description": ""
      }
    },
      "column": "quantity",
      "properties": {
        "dtype": "number",
        "std": 0.44513190725972585,
        "min": 1,
        "max": 4,
        "samples": [
          4,
          1
        ],
        "num_unique_values": 4,
        "semantic_type": "",
        "description": ""
      }
    },
    {
      "column": "order_status",
      "properties": {
        "dtype": "category",
        "samples": [
          "Returned to seller"
        ],
        "num_unique_values": 2,
        "semantic_type": "",
        "description": ""
      }
    }
  ],
  "field_names": [
    "order_no",
    "order_date",
    "buyer",
    "ship_city",
    "ship_state",
    "sku",
    "quantity",
    "order_status"
  ]
}

\end{lstlisting}
}
\section{Participants' Verification Processes}
We include the verification processes of all participants in Table~\ref{tab:all-workflows}.\enlargethispage{24pt}

\section{Example Interface In The User Study}
\begin{figure*}[t]
  \centering
  \includegraphics[width=.92\linewidth]{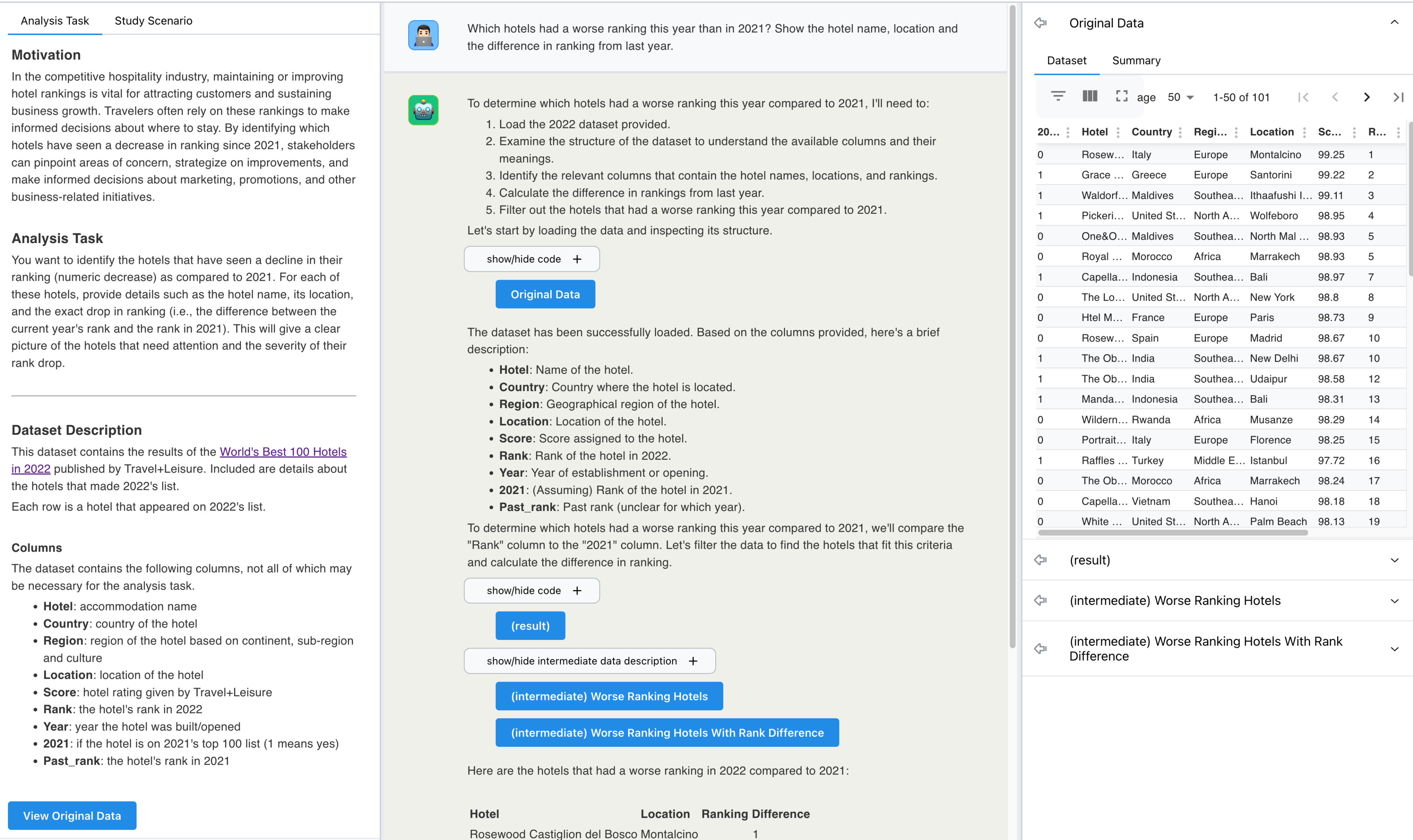}
  \caption{\textbf{An example study task}. We include the description of the analysis goal and dataset in the left side-panel.}
  \Description{}
  \label{fig:interfaceAll}
\end{figure*}

An example of the interface used during the study for T7 is shown in Fig.~\ref{fig:interfaceAll}.


\begin{table*} 
    \small
\begin{tabular}{llllll}
\toprule
Verification Processes & Count & Occurrences \\
    \midrule
    \colorbox{lgray}{\textcolor{tgray}{S}}---\colorbox{lorange}{\textcolor{torange}{PD}}---\colorbox{lgreen}{\textcolor{tgreen}{C}} & 11 & T4 [P10], T6 [P10, P20, P8], T9 [P10, P11, P21], T3 [P2, P22], T8 [P7], T7 [P8]\\
    \vspace{2pt}\colorbox{lgray}{\textcolor{tgray}{S}}---\colorbox{lorange}{\textcolor{torange}{PD}}---\colorbox{lyellow}{\textcolor{tyellow}{N}}---\colorbox{lteal}{\textcolor{tteal}{D}}---\colorbox{lgreen}{\textcolor{tgreen}{C}} & 7 & T1 [P1], T2 [P12, P5], T4 [P2], T8 [P22], T10 [P22], T9 [P6] \\
    \vspace{2pt}\colorbox{lgray}{\textcolor{tgray}{S}}---\colorbox{lorange}{\textcolor{torange}{PD}}---\colorbox{lyellow}{\textcolor{tyellow}{N}}---\colorbox{lgreen}{\textcolor{tgreen}{C}} & 5 & T5 [P10, P14], T2 [P2], T4 [P6], T1 [P7] \\
    \vspace{2pt}\colorbox{lgray}{\textcolor{tgray}{S}}---\colorbox{lorange}{\textcolor{torange}{PD}}---\colorbox{lyellow}{\textcolor{tyellow}{N}}---\colorbox{lorange}{\textcolor{torange}{PD}}---\colorbox{lgreen}{\textcolor{tgreen}{C}} & 5 & T3 [P12, P5], T7 [P12], T10 [P20], T6 [P4] \\
    \vspace{2pt}\colorbox{lgray}{\textcolor{tgray}{S}}---\colorbox{lblue}{\textcolor{tblue}{DP}}---\colorbox{lyellow}{\textcolor{tyellow}{N}}---\colorbox{lteal}{\textcolor{tteal}{D}}---\colorbox{lgreen}{\textcolor{tgreen}{C}} & 3 & T4 [P15], T7 [P16], T9 [P9] \\
    \vspace{2pt}\colorbox{lgray}{\textcolor{tgray}{S}}---\colorbox{lorange}{\textcolor{torange}{PD}}---\colorbox{lyellow}{\textcolor{tyellow}{N}}---\colorbox{lblue}{\textcolor{tblue}{DP}}---\colorbox{lyellow}{\textcolor{tyellow}{N}}---\colorbox{lgreen}{\textcolor{tgreen}{C}} & 2 & T2 [P1], T10 [P13] \\
    \vspace{2pt}\colorbox{lgray}{\textcolor{tgray}{S}}---\colorbox{lred}{\textcolor{tred}{P}}---\colorbox{lyellow}{\textcolor{tyellow}{N}}---\colorbox{lorange}{\textcolor{torange}{PD}}---\colorbox{lgreen}{\textcolor{tgreen}{C}} & 2 & T8 [P11], T5 [P8] \\
    \vspace{2pt}\colorbox{lgray}{\textcolor{tgray}{S}}---\colorbox{lblue}{\textcolor{tblue}{DP}}---\colorbox{lgreen}{\textcolor{tgreen}{C}} & 2 & T5 [P17], T1 [P20] \\
    \vspace{2pt}\colorbox{lgray}{\textcolor{tgray}{S}}---\colorbox{lred}{\textcolor{tred}{P}}---\colorbox{lyellow}{\textcolor{tyellow}{N}}---\colorbox{lred}{\textcolor{tred}{P}}---\colorbox{lgreen}{\textcolor{tgreen}{C}} & 2 & T7 [P4, P6] \\
    \vspace{2pt}\colorbox{lgray}{\textcolor{tgray}{S}}---\colorbox{lblue}{\textcolor{tblue}{DP}}---\colorbox{lyellow}{\textcolor{tyellow}{N}}---\colorbox{lorange}{\textcolor{torange}{PD}}---\colorbox{lgreen}{\textcolor{tgreen}{C}} & 1 & T3 [P1] \\
    \vspace{2pt}\colorbox{lgray}{\textcolor{tgray}{S}}---\colorbox{lorange}{\textcolor{torange}{PD}}---\colorbox{lteal}{\textcolor{tteal}{D}}---\colorbox{lgreen}{\textcolor{tgreen}{C}} & 1 & T9 [P13] \\
    \vspace{2pt}\colorbox{lgray}{\textcolor{tgray}{S}}---\colorbox{lorange}{\textcolor{torange}{PD}}---\colorbox{lyellow}{\textcolor{tyellow}{N}}---\colorbox{lorange}{\textcolor{torange}{PD}}---\colorbox{lyellow}{\textcolor{tyellow}{N}}---\colorbox{lteal}{\textcolor{tteal}{D}}---\colorbox{lgreen}{\textcolor{tgreen}{C}} & 1 & T6 [P14] \\
    \vspace{2pt}\colorbox{lgray}{\textcolor{tgray}{S}}---\colorbox{lteal}{\textcolor{tteal}{D}}---\colorbox{lblue}{\textcolor{tblue}{DP}}---\colorbox{lgreen}{\textcolor{tgreen}{C}} & 1 & T1 [P15] \\
    \vspace{2pt}\colorbox{lgray}{\textcolor{tgray}{S}}---\colorbox{lred}{\textcolor{tred}{P}}---\colorbox{lyellow}{\textcolor{tyellow}{N}}---\colorbox{lblue}{\textcolor{tblue}{DP}}---\colorbox{lgreen}{\textcolor{tgreen}{C}} & 1 & T6 [P17] \\
    \vspace{2pt}\colorbox{lgray}{\textcolor{tgray}{S}}---\colorbox{lorange}{\textcolor{torange}{PD}}--- \colorbox{pgray}{\textcolor{red}{R}} & 1 & T9 [P18] \\
    \vspace{2pt}\colorbox{lgray}{\textcolor{tgray}{S}}---\colorbox{lteal}{\textcolor{tteal}{D}}---\colorbox{lgreen}{\textcolor{tgreen}{C}} & 1 & T8 [P19] \\
    \vspace{2pt}\colorbox{lgray}{\textcolor{tgray}{S}}---\colorbox{lblue}{\textcolor{tblue}{DP}}---\colorbox{lorange}{\textcolor{torange}{PD}}---\colorbox{lyellow}{\textcolor{tyellow}{N}}---\colorbox{lred}{\textcolor{tred}{P}}---\colorbox{lorange}{\textcolor{torange}{PD}}---\colorbox{lgreen}{\textcolor{tgreen}{C}} & 1 & T9 [P19] \\
    \vspace{2pt}\colorbox{lgray}{\textcolor{tgray}{S}}---\colorbox{lorange}{\textcolor{torange}{PD}}---\colorbox{lyellow}{\textcolor{tyellow}{N}}---\colorbox{lblue}{\textcolor{tblue}{DP}}---\colorbox{lgreen}{\textcolor{tgreen}{C}} & 1 & T4 [P21] \\
    \vspace{2pt}\colorbox{lgray}{\textcolor{tgray}{S}}---\colorbox{lteal}{\textcolor{tteal}{D}}---\colorbox{lyellow}{\textcolor{tyellow}{N}}---\colorbox{lblue}{\textcolor{tblue}{DP}}---\colorbox{pgray}{\textcolor{red}{R}} & 1 & T7 [P3] \\
    \vspace{2pt}\colorbox{lgray}{\textcolor{tgray}{S}}---\colorbox{lred}{\textcolor{tred}{P}}---\colorbox{lyellow}{\textcolor{tyellow}{N}}---\colorbox{lorange}{\textcolor{torange}{PD}}---\colorbox{lyellow}{\textcolor{tyellow}{N}}---\colorbox{lteal}{\textcolor{tteal}{D}}---\colorbox{lgreen}{\textcolor{tgreen}{C}} & 1 & T5 [P4] \\
    \vspace{2pt}\colorbox{lgray}{\textcolor{tgray}{S}}---\colorbox{lorange}{\textcolor{torange}{PD}}---\colorbox{pgray}{\textcolor{red}{R}} & 1 & T8 [P5] \\
    \vspace{2pt}\colorbox{lgray}{\textcolor{tgray}{S}}---\colorbox{lteal}{\textcolor{tteal}{D}}---\colorbox{lyellow}{\textcolor{tyellow}{N}}---\colorbox{lorange}{\textcolor{torange}{PD}}---\colorbox{lgreen}{\textcolor{tgreen}{C}} & 1 & T10 [P7] \\
    \bottomrule
\end{tabular}
  \caption{\textbf{All Participants' Behaviors in Their Workflows.} \colorbox{pgray}{\textcolor{red}{R}}~is when the participant ran out of time on the task or was unable to confirm or reject the AI-generated output has an error. In 52 verification workflows, we observed 3 such cases.}
  \label{tab:all-workflows}
\end{table*}

\end{document}